\documentclass[aps,nofootinbib,showpacs]{revtex4-2}  
\usepackage{graphicx,subfigure}
\usepackage{float}
\usepackage{bm}
\usepackage{braket}
\usepackage{amsmath}
\usepackage{amssymb}
\usepackage{amscd}
\usepackage{latexsym}
\usepackage{slashed}
\usepackage{color}
\usepackage{graphicx}
\usepackage[normalem]{ulem}
\usepackage{color}
\definecolor{orange}{cmyk}{0,0.5,1,0}
\usepackage{verbatim}
\usepackage{rotating}
\usepackage{diagbox}
\usepackage{cancel}
\usepackage[utf8]{inputenc}
\usepackage{array}
\usepackage{hyperref}

\def\lsim{\raise0.3ex\hbox{$\;<$\kern-0.75em\raise-1.1ex\hbox{$\sim\;$}}}
\def\gsim{\raise0.3ex\hbox{$\;>$\kern-0.75em\raise-1.1ex\hbox{$\sim\;$}}}

\def\be{\begin{equation}}
\def\ee{\end{equation}}
\def\bea{\begin{eqnarray}}
\def\eea{\end{eqnarray}}
\def\nn{\nonumber}

\begin{document}
\title{Explaining electron and muon $g-2$ anomalies in an  \\[0.15cm] Aligned 2-Higgs Doublet Model with Right-Handed Neutrinos }

\author{Luigi Delle Rose}
\email[]{ldellerose@ifae.es}
\affiliation{\small Institut de Fisica d'Altes Energies (IFAE), The Barcelona Institute of Science and Technology, Campus UAB, 08193 Bellaterra (Barcelona), Spain}
\affiliation{\small School of Physics and Astronomy, University of Southampton,
	Southampton, SO17 1BJ, United Kingdom}

\author{Shaaban Khalil}
\email[]{skhalil@zewailcity.edu.eg}

\affiliation{Center for Fundamental Physics, Zewail City of Science and Technology, 6 October City, Giza 12588, Egypt}

\author{Stefano Moretti}
\email[]{s.moretti@soton.ac.uk}
\affiliation{\small School of Physics and Astronomy, University of Southampton,
	Southampton, SO17 1BJ, United Kingdom}

%\date{\today}
\begin{abstract}
We explain anomalies currently present in various data samples used for the measurement of the anomalous magnetic moment of electron ($a_e$) and muon ($a_\mu$) in terms of an Aligned 2-Higgs Doublet Model with right-handed neutrinos. The explanation is driven by one and two-loop topologies wherein a very light CP-odd neutral Higgs state ($A$) contributes significantly to $a_\mu$ but negligibly to $a_e$, so as to revert the sign of the new physics corrections in the former case with respect to the latter, wherein the dominant contribution is due to a charged Higgs boson ($H^\pm$) and heavy neutrinos with mass at the electroweak scale. For the region of parameter space of our new physics model which explains the aforementioned anomalies we also predict an almost  background-free smoking-gun signature of it,  consisting  of $H^\pm A$ production followed by Higgs boson decays yielding multi-$\tau$ final states, which can be pursued at the Large Hadron Collider. 

%We explain  anomalies currently present in various data samples used for the measurement of the anomalous magnetic moment of  electron ($a_e$) and muon 
%($a_\mu$) in terms of an Aligned 2-Higgs Doublet Model with right-handed neutrinos. The explanation is driven by one-loop topologies wherein a very light CP-odd neutral Higgs state, $A$, with mass between 10 and 60 GeV, contributes significantly to $a_\mu$ but negligibly to $a_e$, so as to revert the sign of the new physics corrections  in the former case with respect to the latter, wherein the dominant contribution is due to a charged Higgs boson with $m_{H^\pm}\sim m_t$, entering one-loop diagrams also involving heavy neutrinos with mass at the electroweak scale. For the region of parameter space of our new physics model which explains the aforementioned anomalies we also predict an almost  background-free
%smoking-gun signature of it,  consisting  of $H^\pm A$ production followed by Higgs boson decays yielding multi-$\tau$ final states, which can be pursued at the Large Hadron Collider. 

\end{abstract}
\maketitle
%%%%%%%%%%%%%%%%%%%%%%%%%
\section{Introduction}

It is tempting to conclude that the time-honoured discrepancy between the Standard Model (SM) prediction for the muon anomalous magnetic moment and its experimental measurement is a firm indication of New Physics (NP) Beyond the SM (BSM). Moreover, after improving the determination of the fine structure constant, it recently turned out that there is also a significant difference between the experimental result of the electron anomalous magnetic moment and the corresponding SM prediction. According to the latest results, we have the following deviations in the anomalous magnetic moments of muon and electrons \cite{Keshavarzi:2020bfy,Parker:2018vye}: 
\bea
\label{eq:g2mu}
\delta a_{\mu} &=& a_{\mu}^{\rm exp} - a_{\mu}^{\rm SM} = (278\pm 88) \times 10^{-11} \,, \nn \\
\delta a_{e} &=& a_{e}^{\rm exp} - a_{e}^{\rm SM} = (-87\pm 36) \times 10^{-14},
\eea
which indicate a $3.1 \sigma$ and $2.4\sigma$ discrepancy between theory and experiment, respectively. Fermilab and J-PARC experiments 
\cite{Semertzidis:1999kv,Farley:2003wt} are going to explore these anomalies in the near future with much higher precision, but now it is worthwhile speculating what possible NP phenomena might lie behind these two measurements. In doing so, it should be noted that $\delta a_{e}$ and $\delta a_{\mu}$ have opposite signs, which provides a challenge for any BSM explanation attempting to account for both of them simultaneously. This generated growing interest and several extensions of the SM have been analysed as possible origin of the results in (\ref{eq:g2mu}). 

It is clear that any Electro-Weak (EW) scale NP effects that may explain the $a_{\mu}$ result will lead to corrections to $a_e$ of order $10^{-5}$ times smaller, due to the typical relative suppression generated by the mass ratio $(m_e/m_\mu)^2$, and, crucially, with the same sign. Therefore, the anomalies of $a_\mu$ and $a_e$ cannot be resolved simultaneously with the same NP contribution, unless it violates lepton flavour universality in a very peculiar way, so as to give a positive contribution to $a_\mu$ and a negative one to $a_e$. Some attempts along this line were in fact pursued by Ref. \cite{Liu:2018xkx,Han:2018znu,Endo:2019bcj,Bauer:2019gfk,Badziak:2019gaf,CarcamoHernandez:2020pxw,Haba:2020gkr,Bigaran:2020jil,Calibbi:2020emz,Chen:2020jvl,Jana:2020pxx,Li:2020dbg,Chun:2020uzw,Jana:2020joi,Arbelaez:2020rbq,DelleRose:2020qak,Crivellin:2018qmi,Dutta:2020scq,Hati:2020fzp,CarcamoHernandez:2019ydc,Crivellin:2019mvj,Botella:2020xzf}.     

In this paper, we analyse the anomalous magnetic moment of muon and electron in a 2HDM with RH neutrinos and aligned Yukawa couplings. We emphasise that, in this class of models, one can account for the $a_e$ through one-loop effects generated by the exchange of RH neutrinos and charged Higgs bosons. At the same time, the measured value of $a_\mu$ can be obtained accurately through two-loop effects generated by a light CP-odd neutral Higgs state in combination with charged leptons. This phenomenology requires the $H^\pm$ and $A$ states to be relatively light, so that their pair production process has a sizeable cross section at the Large Hadron Collider (LHC), thereby enabling one to fingerprint this A2HDM with RH neutrinos in the years to come.

The plan of this paper is as follows. In the next section we describe our NP scenario. In the following one we present the formulae for $a_e$ and $a_\mu$. After this, we present our results for the two anomalous magnetic moments and the aforementioned $H^\pm A$ signature in two separate subsections. We then conclude.

%%%%%%%%%%%%%%%%%%%%%%%%%

\section{A2HDM with RH Neutrinos}

The most general Yukawa Lagrangian of the 2HDM can be written as
\bea
\label{eq:yukL}
- \mathcal L_Y &=&   \bar Q_L' \left( Y'_{1d}  \Phi_1 + Y'_{2d}  \Phi_2 \right) d_R' + \bar Q_L' \left( Y'_{1u} \tilde \Phi_1 + Y'_{2u} \tilde \Phi_2 \right) u_R'  
+ \bar L'_L \left( Y'_{1\ell}  \Phi_1 + Y'_{2\ell}  \Phi_2 \right) \ell_R'   \nn \\
&+&  \bar L'_L \left( Y'_{1\nu}  \tilde \Phi_1 +  Y'_{2\nu} \tilde \Phi_2 \right) \nu_R'  + \textrm{h.c.},
\eea
where the quark $Q_L', u_R', d_R'$ and lepton $L_R', \ell_R', \nu_R'$ fields are defined in the weak interaction basis and we also included the couplings of the Left-Handed (LH) lepton doublets with the RH neutrinos. The $\Phi_{1,2}$ fields are the two Higgs doublets in the Higgs basis and, as customary, $\tilde \Phi_i = i \sigma^2 \Phi_i^*$.
The Yukawa couplings $Y_{1j}'$ and $Y_{2j}'$, with $j = u,d,\ell$, are $3\times 3$ complex matrices while $Y_{1\nu}'$ and $Y_{2\nu}'$ are $3 \times n_R$ matrices, with $n_R$ being the number of RH neutrinos.
Besides implementing the standard $Z_2$ symmetry, potentially dangerous tree-level Flavour Changing Neutral Currents (FCNCs) can be tamed by requiring the alignment in flavour space of the two Yukawa matrices that couple to the same right-handed quark or lepton. This implies\footnote{We have assumed real $\zeta_f$. Notice also that the alignment in the neutrino sector is not a a consequence of the requirement of the absence of FCNCs. Nevertheless, we assume that the same mechanism that provides the alignment in the SM flavour space also holds for neutrinos. }
\bea
Y'_{2,d} = \zeta_d Y'_{1,d} \equiv \zeta_d Y'_d, 	\qquad    Y'_{2,u} = \zeta_u Y'_{1,u} \equiv \zeta_u Y'_u,  \qquad Y'_{2,\ell} = \zeta_\ell Y'_{1,\ell} \equiv \zeta_\ell Y'_\ell, \qquad  Y'_{2,\nu} = \zeta_\nu Y'_{1,\nu} \equiv \zeta_\nu Y'_\nu \,.
\eea
Renormalisation group effects can introduce some misalignment in the Yukawa couplings. These provide negligible FCNC contributions in the quark sector  suppressed by mass hierarchies $m_q m_{q'}^2/v^3$ \cite{Jung:2010ik,Li:2014fea}.

\begin{table}[h]
\centering
\begin{tabular}{|ccccc|}
\hline
Aligned & Type I & Type II & Type III & Type IV \\
\hline \hline
$\zeta_u$ & $\cot \beta$ & $\cot \beta$ & $\cot \beta$ & $\cot \beta$ \\
$\zeta_d$ & $\cot \beta$ & $- \tan \beta$ & $-\tan \beta$ & $\cot \beta$ \\
$\zeta_l$ & $\cot \beta$ & $- \tan \beta$ & $\cot \beta$ & $-\tan \beta$ \\
\hline
\end{tabular}
\caption{Relation between the $\zeta_f$ couplings of the A2HDM and the ones of the $Z_2$ symmetric scenarios. \label{tab:2hdms}}
\end{table}

The Yukawa Lagrangian in Eq.~(\ref{eq:yukL}) generates a Dirac mass matrix for the standard neutrinos and can also be supplemented by a Majorana mass term $M_R'$ for the RH ones
\bea
- \mathcal L_{M_R} = \frac{1}{2} \nu_R'^{\, T} C M_R' \nu_R' + \textrm{h.c.},
\eea
where $C$ is the charge-conjugation operator. 
In particular, by exploiting a bi-unitary transformation in the charged lepton sector and a unitary transformation on the RH neutrinos, $L_L' = U_L \, L_L, \, \ell'_R = U_R^\ell \, \ell_R$ and $\nu'_R = U_R^\nu \, \nu_R$, it 
is always possible to diagonalise (with real eigenvalues) the charged lepton and Majorana mass matrices at the same time,
\bea
U_L^\dag Y'_{\ell} U_R^e &=& Y_{\ell} \equiv \frac{\sqrt{2}}{v} \textrm{diag} (m_e, m_\mu, m_\tau) \,, \nn \\
U_R^{^\nu \, T} M'_R U_R^\nu &=& M_R \equiv \textrm{diag}( M_1 , \ldots M_{n_R} ),
\eea
while $Y_{\nu} =  U_L^\dag Y'_{\nu} U_R^\nu$ remains non-diagonal. 
In this basis the neutrino mass matrix can be written as
\bea
\label{eq:mass_matrix}
- \mathcal L_{\mathcal M_\nu} = \frac{1}{2} N_L^T C \mathcal M N_L  + \textrm{h.c.}  = \frac{1}{2} (\nu_L^T \, \nu_R^{c \,\, T}) C  \left( \begin{array}{cc} 0 & M_D \\ M_D^T & M_R \end{array} \right)  \left( \begin{array}{c} \nu_L \\ \nu_R^c \end{array} \right),
\eea
with $M_D = \frac{v}{\sqrt{2}} Y_{\nu}^*$ being the neutrino Dirac mass.
This can be diagonalised with a unitary $(3 + n_R) \times (3 + n_R)$ matrix $U$, via 
\bea
\left( \begin{array}{c} \nu_L \\ \nu_R^c \end{array} \right) = U \left( \begin{array}{c} \nu_l \\ \nu_h \end{array} \right) \equiv \left( \begin{array}{cc} U_{Ll} & U_{Lh} \\ U_{R^c l} & U_{R^c h} \end{array} \right) \left( \begin{array}{c} \nu_l \\ \nu_h \end{array} \right) ,
\eea
such that $\mathcal M_\nu = U^T \mathcal M U$ provides the masses of the three light active neutrinos $\nu_l$ and of the remaining $n_R$ heavy sterile neutrinos $\nu_h$.

The Yukawa interactions of the physical (pseudo)scalars\footnote{Note that, in a generic 2HDM with complex Higgs doublet fields, of the initial 8 degrees of freedom, upon EW Symmetry Breaking (EWSB), 5 survive as physical Higgs states: 2 CP-even, $h$ and $H$ (with, conventionally, $m_h<m_H$), 1 CP-odd, $A$, and 2 charged ones with undefined CP, $H^\pm$.} with the mass eigenstate fermions are then described by
\bea
- \mathcal L_Y &=&  \frac{\sqrt{2}}{v} \bigg[   \bar u  ( - \zeta_u \, m_u \, V_{ud} \, P_L + \zeta_d \, V_{ud} \, m_d \, P_R  ) d   
+ \bar \nu_l  ( - \zeta_\nu \, m_{\nu_l} \, U^\dag_{L l}  \,  P_L   + \zeta_\ell  \,  U^\dag_{L l}  \, m_\ell \, P_R )  \ell       \nn \\
&+&  \bar \nu_h  ( - \zeta_\nu \, m_{\nu_h} \, U^\dag_{L h}  \,  P_L   + \zeta_\ell  \,  U^\dag_{L h}  \, m_\ell \, P_R )  \ell   \bigg]  H^+  + \textrm{h.c.}  \nn \\
&+& \frac{1}{v} \sum_{\phi=h,H,A} \sum_{f=u,d,\ell} \xi_f^\phi \, \phi \, \bar f  \, m_f \, P_R \,  f   
+ \frac{1}{v} \sum_{\phi=h,H,A} \xi_\nu^\phi \, \phi (\bar \nu_l \, U_{Ll}^\dag + \bar \nu_h \, U_{Lh}^\dag) P_R (U_{Ll} \, m_{\nu_l} \, \nu_l^c + U_{Lh} \, m_{\nu_h} \, \nu_h^c) + \textrm{h.c.}, 
\eea
where the couplings of the neutral Higgs states to the fermions are given by 
\bea
\xi_{u, \nu}^\phi = \mathcal R_{i1} + ( \mathcal R_{i2} - i  \mathcal R_{i3} ) \zeta_u^*   \,, \qquad
\xi_{d,\ell}^\phi = \mathcal R_{i1} + ( \mathcal R_{i2} + i  \mathcal R_{i3} ) \zeta_{d,\ell}, 
\eea
where the matrix $\mathcal R$ diagonalises the scalar mass matrix.
Because of the alignment of the Yukawa matrices all the couplings of the (pseudo)scalar fields to fermions are proportional to the corresponding mass matrices, hence the A2HDM acronym. Therefore, this 2HDM realisation is notably different from the standard four Types \cite{Gunion:1989we,Gunion:1992hs,Branco:2011iw}, wherein the Yukawa couplings are fixed to well defined functions of the ratio of the Vacuum Expectation Values (VEVs) of the two Higgs doublets, denoted by $\tan\beta$, see Tab.~\ref{tab:2hdms}.

Then, the charged Higgs boson currents in the lepton sector are given by:
\bea
- \mathcal L_{Y}^\textrm{CC} = \frac{\sqrt{2}}{v} \zeta_\ell \left[ (\bar \nu_l \, U^\dag_{L l}  +   \bar \nu_h \, U^\dag_{Lh}) m_\ell  \, P_R \, \ell \right]  H^+  
-  \frac{\sqrt{2}}{v}  \zeta_\nu \left[ (\bar \nu_l \, U^\dag_{L l}  \, m_{\nu_l} +   \bar \nu_h \, U^\dag_{Lh} \, m_{\nu_h})  \, P_L \, \ell \right]   H^+   + \textrm{h.c.}
\eea
Finally, the  neutral and charged gauge boson interactions of the neutrinos are
\bea
 \mathcal L_Z &=& \frac{g}{2 \cos \theta_W}  (\bar \nu_l \, U_{Ll}^\dag + \bar \nu_h \, U_{Lh}^\dag) \gamma^\mu  (U_{Ll} \, \nu_l  +  U_{Lh} \, \nu_h )  Z_\mu, \nn \\
 \mathcal L_W &=& - \frac{g}{\sqrt{2}} \left[  (\bar \nu_l \, U^\dag_{L l} + \bar \nu_h \, U^\dag_{L h}) \gamma^\mu P_L \, \ell \right] W^{+}_\mu  + \textrm{h.c.}
\eea
We refer to \cite{DelleRose:2019ukt} for further details on the model.

%%%%%%%%%%%%%%%%%%%%%%%%%%%%%%%
\section{Anomalous magnetic moments}

\begin{figure}
\subfigure[]{\includegraphics[scale=0.35]{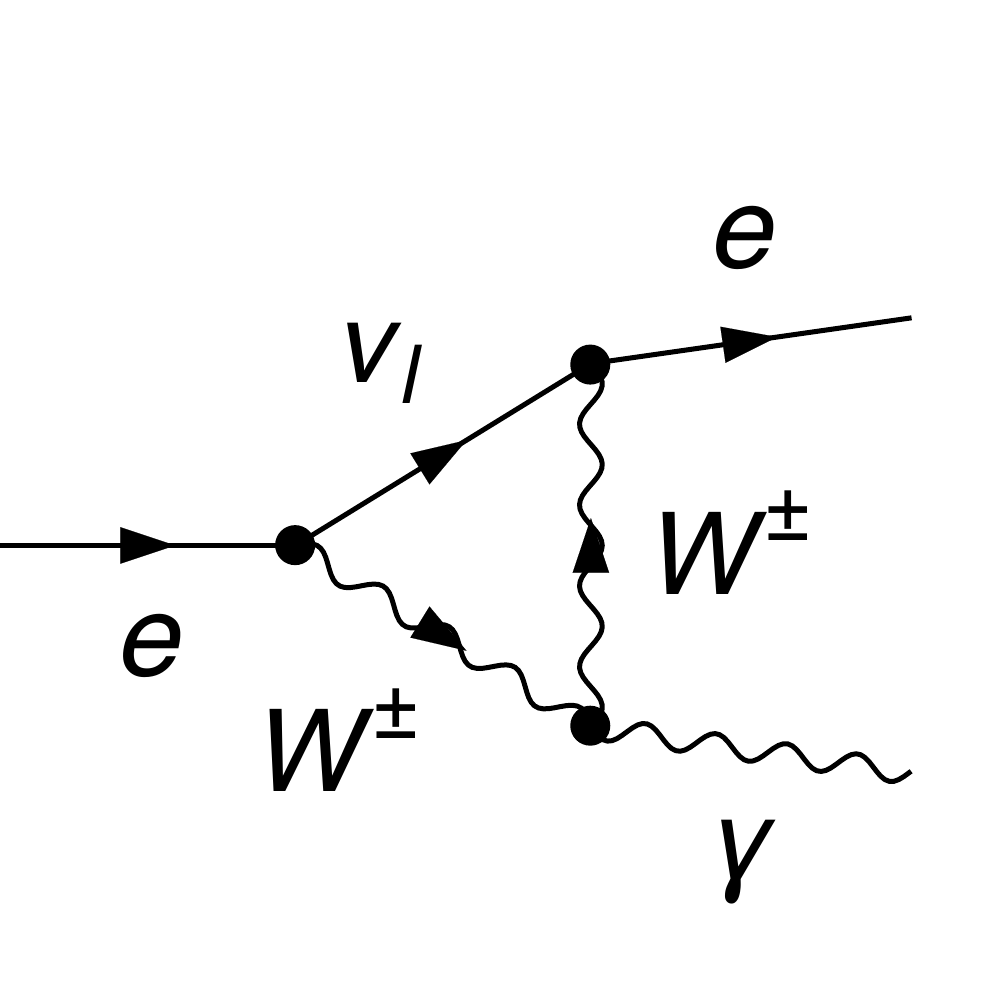}}
\subfigure[]{\includegraphics[scale=0.35]{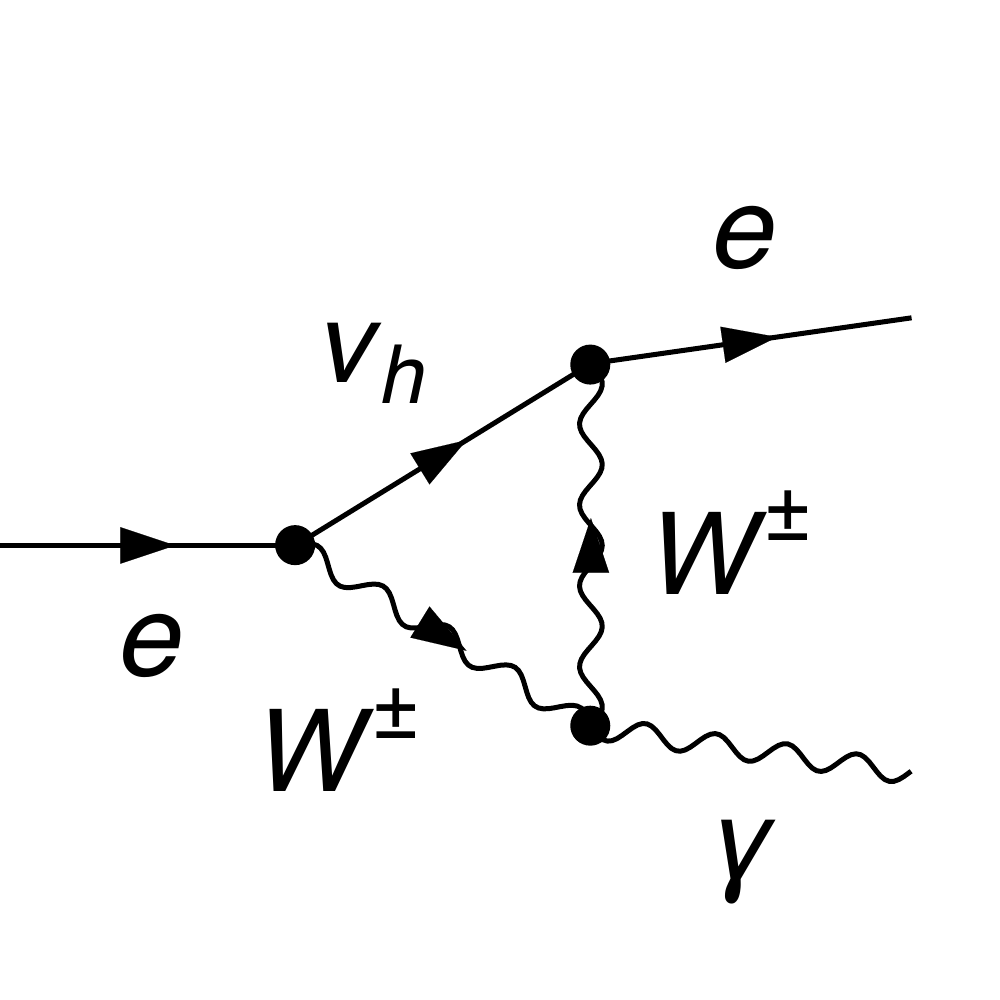}}
\subfigure[]{\includegraphics[scale=0.35]{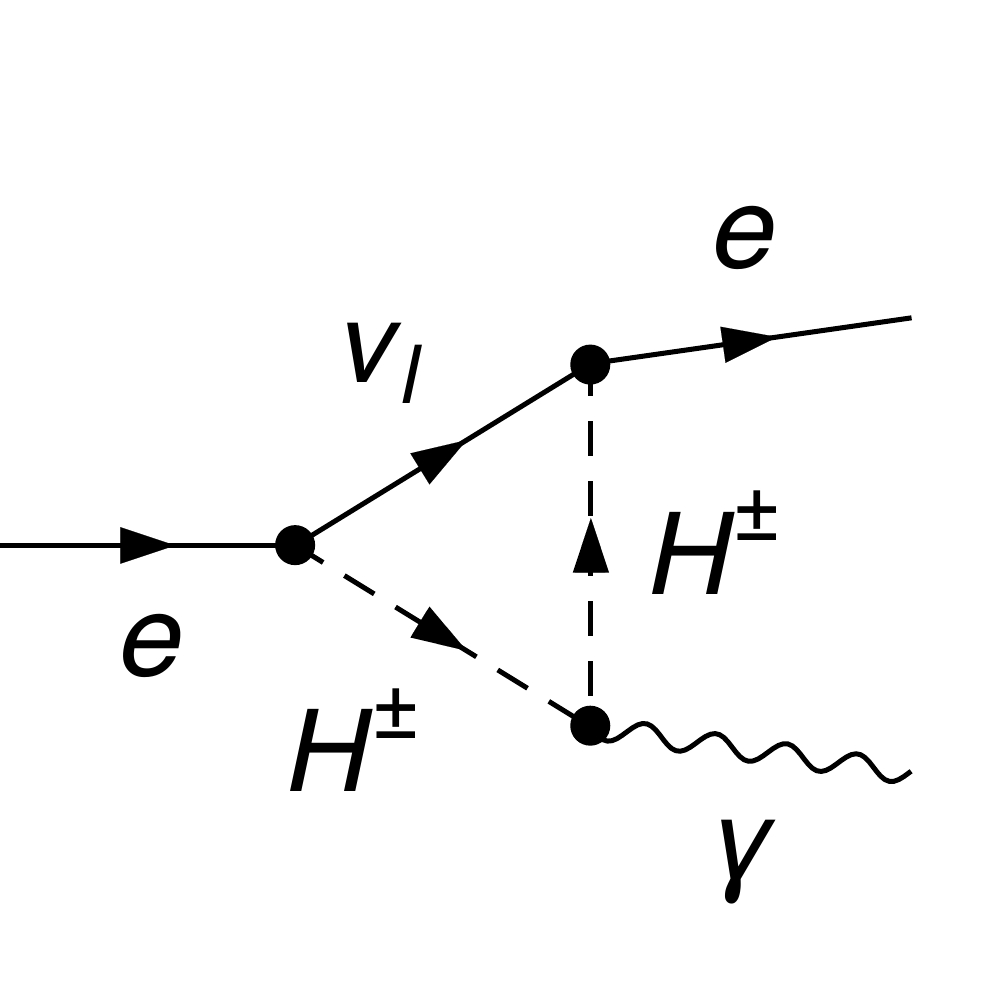}}
\subfigure[]{\includegraphics[scale=0.35]{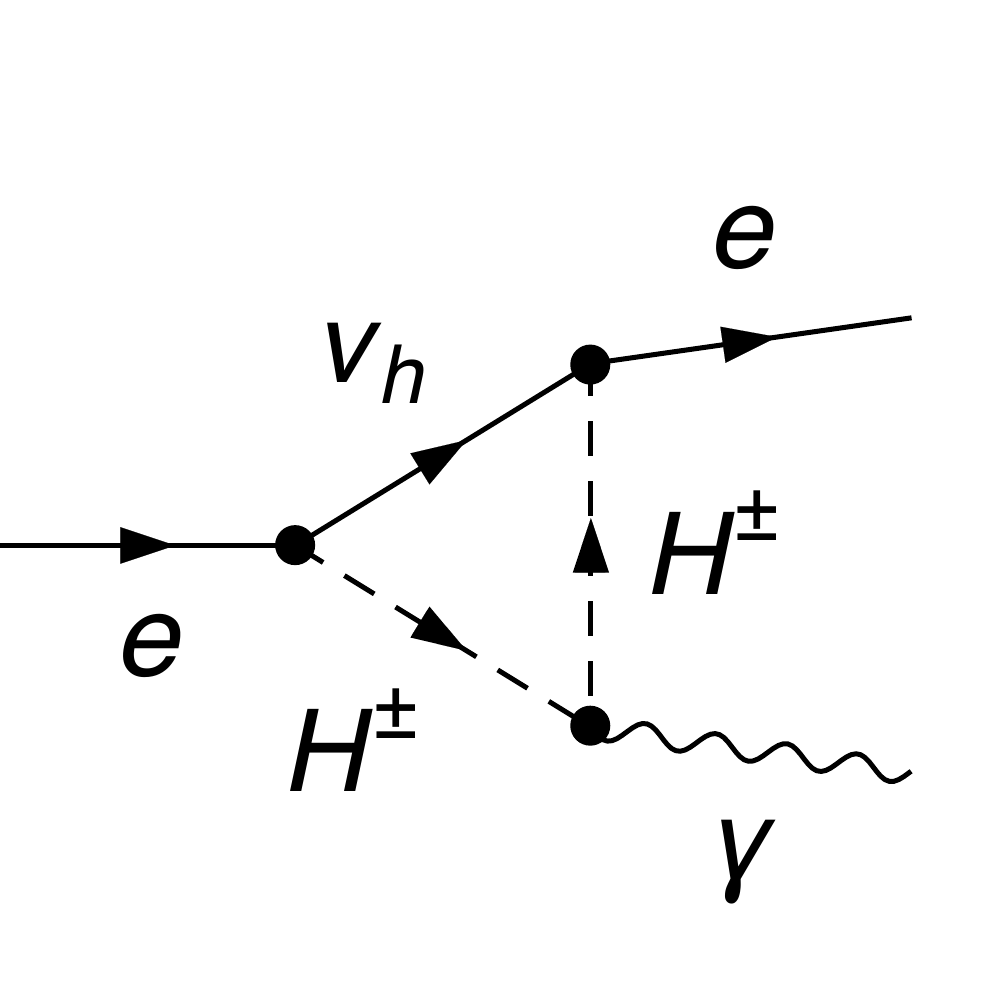}}
\caption{Relevant Feynman diagrams contributing to the $g-2$ of the electron at one-loop order. Only the charges vector ($W^\pm$) and charged Higgs ($H^\pm$)  currents are shown.   \label{fig:diagrams}}
\end{figure}

The one-loop contributions to the anomalous magnetic moment of either lepton are
\bea
a_\ell = \frac{G_F \, m_\ell^2}{4 \sqrt{2} \pi^2}  \left[  g_{(a)} + g_{(b)} + g_{(c)} + g_{(d)}  + g_{\textrm{2HDM}}  \right],
\eea
where the individual terms are
\bea
g_{(a)} &=& 2 \sum_{i = 1}^{3} |(U_{Ll})_{\ell \, i}|^2   \left[ \frac{5}{6}  +    \frac{1}{6} \frac{m_\ell^2}{M_W^2}  \right] + \mathcal O(m_\ell^4) \,, \nn \\
g_{(b)} &=&  2 \sum_{i = 1}^{n_{R}} |(U_{Lh})_{\ell \, i}|^2  \left[ \frac{5}{6}  +   \mathcal G_{W^\pm} \left( \frac{m_{\nu_{h_i}}^2}{M_W^2} \right) \right]   + \mathcal O(m_\ell^2)     \,, \nn \\
g_{(c)} &=&  2 \sum_{i = 1}^{3} |(U_{Ll})_{\ell \, i}|^2   \left[    -\frac{\zeta_\ell^2}{12} \frac{m_\ell^2}{M_{H^\pm}^2}    \right]               +    \mathcal O(m_\ell^4) \,, \nn \\
g_{(d)} &=&  2 \sum_{i = 1}^{n_{R}} |(U_{Lh})_{\ell \, i}|^2  \, \mathcal G_{H^\pm} \left( \frac{m_{\nu_{h_i}}^2}{M_{H^\pm}^2} \right)   + \mathcal O(m_\ell^2)  \,, \nn \\
g_\textrm{2HDM} &=&  \mathcal O(m_\ell^2),
\eea
with
\bea
\label{eq:Gfuncs}
\mathcal G_{W^\pm}(x) &=& \frac{-x + 6 x^2 - 3 x^3 - 2 x^4 + 6 x^3 \log x}{4(x - 1)^4}, \nn \\
\mathcal G_{H^\pm}(x) &=& \frac{\zeta_\nu^2}{3} \mathcal G_{W^\pm}(x) + \zeta_\nu \zeta_l \frac{ x (-1 + x^2 - 2 x \log x) }{2 (x - 1)^3} \,.
\eea
The index of the contributions corresponds to the different subfigures in Fig.~\ref{fig:diagrams} where, for simplicity, we show only the diagrams determined by the charged currents.
The contribution $g_{(a)}$ alone would exactly correspond to the SM case if it were not for the rescaling induced by the neutrino mixing matrix.
Nevertheless, the constant terms in $g_{(a)}$ and $g_{(b)}$ sums up to the SM result of $5/3$ due to the unitarity of such a mixing matrix. Therefore, these can be neglected since they do not contribute to the NP part.
The term $g_\textrm{2HDM}$ contains all the  neutral Higgs boson contributions which are typical of the 2HDM alone. These  are typically suppressed by a factor of $m_\ell^2/m_\phi^2$, with $\phi$ being one of the  neutral (pseudo)scalar states of the 2HDM. 

We can then write 
the contribution to $(g-2)_\ell$, $\ell = e, \mu$, due to charged currents as follows:
\be
a_\ell^\pm=a_\ell^{W^\pm} +  
a_\ell^{H^\pm} = \frac{G_F \, m_\ell^2}{2 \sqrt{2} \pi^2} \sum_{i = 1}^{n_{R}} |(U_{Lh})_{\ell \, i}|^2  \left[   \mathcal G_{W^\pm} \left( \frac{m_{\nu_{h_i}}^2}{M_W^2} \right) +   \mathcal G_{H^\pm} \left( \frac{m_{\nu_{h_i}}^2}{M_{H^\pm}^2} \right)  \right]. 
\label{eq:AMM}
\ee

The contribution to $(g-2)_\ell$, $\ell=e,\mu$, from the neutral (pseudo)scalars is
\bea a_\ell^0=
\sum_{\phi=h,H,A} a_\ell^\phi = \frac{G_F \, m_\ell^2}{4 \sqrt{2} \pi^2} \sum_{\phi = h, H , A} (\xi^\phi_\ell)^2   \frac{m_\ell^2}{m_\phi^2} \mathcal F_\phi \left(  \frac{m_\ell^2}{m_\phi^2} \right),
\eea
where 
\bea
{\cal F}_h(x) = F_H(x) \simeq - \frac{7}{6} - \log x \,, \qquad \qquad 
{\cal F}_A(x) \simeq \frac{11}{6} + \log x. 
\eea
For the sake of completeness, we also give the Barr-Zee two-loop diagram contributions, \cite{Barr:1990vd,Czarnecki:1995wq,Chang:1990sf,Cheung:2001hz,Cheung:2009fc,Cherchiglia:2016eui}
\bea
a_\ell^\textrm{two-loop} = \frac{G_F m_\ell^2 \alpha}{4 \sqrt{2} \pi^3 } \sum_{\phi= h,H,A} \sum_f N_f^c Q_f^2 \xi_\ell^\phi \xi_f^\phi \frac{m_\ell^2}{m_\phi^2}  G_\phi \left(\frac{m_\ell^2}{m_\phi^2} \right),
\eea
where $N_c^f$ is the number of colours and $Q_f$ the electric charge while
\bea
G_\phi(x) = \int_0^1 d z \frac{\tilde g_\phi(z)}{z(1-z)-x} \log \frac{z(1-z)}{x} \,, \qquad \textrm{with} \quad \tilde g_\phi(z) = \left\{ \begin{array}{ll} 2 z(1-z) - 1, & \phi = h,H \\ 1, & \phi= A\end{array} \right. \,.
\eea 

The total contribution to the $g-2$ is thus given by $a_\ell = a_\ell^\pm + a_\ell^0 + a_\ell^\textrm{two-loop}$.

Finally we present the Branching Ratio (BR) of the Lepton Flavour Violating (LFV) decays $\ell_\alpha \to \ell_\beta \gamma$ (with $\alpha,\beta=e,\mu,\tau$), as follows: 
\be 
\label{eq:BRltolga}
\textrm{BR}(\ell_\alpha \to \ell_\beta \gamma) =  \mathcal C  \left| \sum_{i = 1}^{n_R}  (U^*_{Lh})_{\alpha i} (U_{L h})_{\beta i} \left[ \mathcal G_{W^\pm} \left( \frac{m_{\nu_{h_i}}^2}{M_W^2} \right)   +   \mathcal G_{H^\pm} \left( \frac{m_{\nu_{h_i}}^2}{M_{H^\pm}^2} \right)  \right] \right|^2,
\ee
with
\bea
\mathcal C = \frac{\alpha_W^3 s_W^2}{256 \pi^2} \left( \frac{m_{\ell_\alpha}}{M_W} \right)^4 \frac{m_{\ell_\alpha}}{\Gamma_{\ell_\alpha}},
\eea
where $\Gamma_{\ell_\alpha}$ is the total decay width of the lepton $\ell_\alpha$ and the loop functions are given above. The structure of the loop corrections is obviously the same as the one appearing above in the charged current corrections to  $(g-2)_\ell$. The measured BR of these  LFV decays will act as a constraint in our analysis.

\section{Results}

The solution of the $a_\mu$ anomaly relies upon a light pseudoscalar state $A$ contributing to the dominant two-loop Barr-Zee diagrams, as customary in 2HDMs.
The explanation of the anomaly is particularly simple in the `lepton-specific' 2HDM scenario, also dubbed Type-IV, in which the couplings of the $A$ and  $H^\pm$ bosons to the leptons can be enhanced (for large $\tan \beta$) while those to the quarks are suppressed (being proportional to $\tan^{-1} \beta$). 
Indeed, while it is always possibile to enhance the couplings to the leptons in any of the four standard realisations of the 2HDM, in Type-I and -III this is done at the cost of increasing the couplings to the up quark (for small $\tan \beta$). As a consequence, one faces a strong constraint from the perturbativity of the top-quark Yukawa coupling. In the Type-II, instead, the couplings to the down-quarks are enhanced (for large $\tan \beta$) and severe bounds are imposed  by flavour physics and direct searches for  extra Higgs bosons.
These issues can be much more easily addressed in the A2HDM since the couplings to leptons and quarks are disentangled and $\zeta_\ell$ can be raised independently of $\zeta_u$ and $\zeta_d$.

It is worth emphasising that a simultaneous explanation of both the $a_e$ and $a_\mu$ anomalies cannot be achieved neither in the $Z_2$ symmetric scenarios of the 2HDM nor in the pure A2HDM, since the contributions to the anomalous moments have a fixed sign as they both originate from the same $\zeta_\ell$. In \cite{Botella:2020xzf}, this constraint has been overcome by decoupling the electron and muon sectors, where all Yukawa matrices can be made diagonal in the fermion mass basis \cite{Penuelas:2017ikk,Botella:2018gzy}.
Here, instead, the degeneracy will be broken by exploiting the lepton non-universality that naturally arises in RH neutrino models: augmenting the A2HDM with RH neutrinos can allow for an independent solution to $a_e$. This is obtained with the one-loop diagrams shown in Fig.~\ref{fig:diagrams} provided that the charged Higgs boson  is not too heavy to suppress the loop corrections.

The mass of the charged Higgs boson  is bounded from below by direct searches at LEP II. In particular, searches for $H^{\pm}$ pair production provide $m_{H^\pm} \gtrsim 93.5$ GeV at 95 \% Confidence Level (CL) \cite{Abbiendi:2013hk} assuming the charged Higgs only decays leptonically into $\tau \nu$. 
Since the mass of the pseudoscalar $A$ state is thus required to be much lighter than the charged one, our scenarios realises the mass hierarchy $m_A \ll m_{H^\pm} \simeq m_H$.
The almost degeneracy between the heavy neutral scalar and the charged Higgs state is induced by the constraints on the EW Precision Observables (EWPOs), i.e., $S, T$ and $U$. Indeed, the most stringent one arises from custodial symmetry and reads as\footnote{The expression for $\Delta T$ assumes the mass hierarchy $m_A \ll m_Z \ll m_{H^\pm} \simeq m_H$ and $\sin(\beta - \alpha) \simeq 1$. }
\bea
\Delta T \simeq \frac{m_H}{32 \pi^2 \alpha v^2} (m_{H^\pm} - m_H),
\eea 
which fixes the mass splitting to $(m_{H^\pm} - m_H) \sim \mathcal O(10$ GeV).

As quoted above, the scenarios with light scalar states is strongly constrained by flavour physics, in particular by neutral meson mixings ($\Delta M_q$ and $\epsilon_K$), leptonic decays of neutral and charged mesons as well as  radiative $B$ decays ($b \to s \gamma$). These mostly depend on $m_{H^\pm}$, $\zeta_{u,d}$. Such measurements are  reconciled in our setup simply by requiring a sufficiently small $\zeta_{u,d}$ which we will set to zero for the sake of simplicity. This in turn implies that the Yukawa interactions in our BSM scenario are purely leptophilic. This configuration also naturally complies with void searches for extra (pseudo)scalars at the LHC. In this respect, we have required that the  Higgs sector of our model is compliant with the experimental constraints implemented in HiggsSignals \cite{Bechtle:2013xfa} (capturing the LHC measurements of the discovered Higgs boson\footnote{In our BSM scenario this is the $h$ state.}) and in HiggsBounds \cite{Bechtle:2020pkv} (enforcing limits following the aforementioned void searches for the $H,A$ and $H^\pm$ states at past and present colliders).

Contributions mediated by the charged Higgs states also affect the leptonic decays $\ell_i \to \ell_j \nu \bar \nu$ at tree level, with the stronger constraint coming  from $\tau \to \mu \nu \bar \nu$ \cite{Kuno:1999jp,Abe:2015oca}. The corresponding bound projects onto the ratio $z = \zeta_\ell^2 \, m_{\tau} m_{\mu}/ m_{H^\pm}^2$ and gives $|z| < 0.72$ at 95\% CL \cite{Zyla:2020zbs}.

Finally, upper bounds on LFV processes, ($\textrm{BR}(\mu \to e \gamma) \le 4.2 \times 10^{-13} \,,  ~\textrm{BR}(\tau \to e \gamma) \le 3.3 \times 10^{-8} \,,  ~\textrm{BR}(\tau \to \mu \gamma) \le 4.4 \times 10^{-8}$ at 90\% CL) constrain the RH neutrinos interactions with the charged leptons. The charged Higgs boson also affects these decays with a large contribution. 
Since a RH neutrino is only employed in the explanation of the $a_e$ anomaly, a non-negligible mixing is strictly required with the electron family. Therefore, the stringent constraint from $\mu \to e \gamma$ and the milder one from $\tau \to e \gamma$ can be satisfied by simply relying on the hierarchy $|(U_{Lh})_{\tau \, \nu_h}|, |(U_{Lh})_{\mu \, \nu_h}| \ll |(U_{Lh})_{e \, \nu_h}|$.

\subsection{Predictions for $\delta a_e$ and $\delta a_\mu$}

The contribution to $\delta a_e$ arising from the $W^\pm$,  encoded in the $ \mathcal G_{W^\pm}$ function defined in Eq.~(\ref{eq:Gfuncs}), is negative but it can never be enhanced being fixed by the gauge interactions. For $m_{\nu_{h_i}}^2/M_W^2 \gg 1$, $\mathcal G_{W^\pm} \simeq - 1/2$. The impact of the charged Higgs boson in the loop functions is, however, much different. As an example, for large heavy neutrino masses, it saturates to $\mathcal G_{H^\pm} \simeq \zeta_\ell \zeta_\nu/2 - \zeta_\nu^2/6$ or behaves as $\mathcal G_{H^\pm} \simeq (\zeta_\ell \zeta_\nu/2 - \zeta_\nu^2/12) (m_{\nu_{h}}^2/m_{H^\pm}^2)$ for larger $m_{H^\pm}$. In both cases, the solution of the $a_e$ anomaly is facilitated by large and opposite $\zeta_\ell$ and $\zeta_\nu$. The same effect would also push the predicted $a_\mu$ in the opposite direction with respect to the current measurement. This is not an issue since the same hierarchy $|(U_{Lh})_{\mu \, \nu_h}| \ll |(U_{Lh})_{e \, \nu_h}|$ required to prevent the LFV bounds also suppresses the contribution of the charged Higgs boson  to the muon $g-2$. 
As well known in the literature, the latter can be explained in the 2HDM by the two-loop Barr-Zee diagrams of the neutral scalars which provide a positive correction for sufficiently light $A$. 
This contribution may compete in $a_e$ against the one-loop effects discussed above but it is found to be subdominant in most of the parameter space.

\begin{figure}
\includegraphics[scale=0.45]{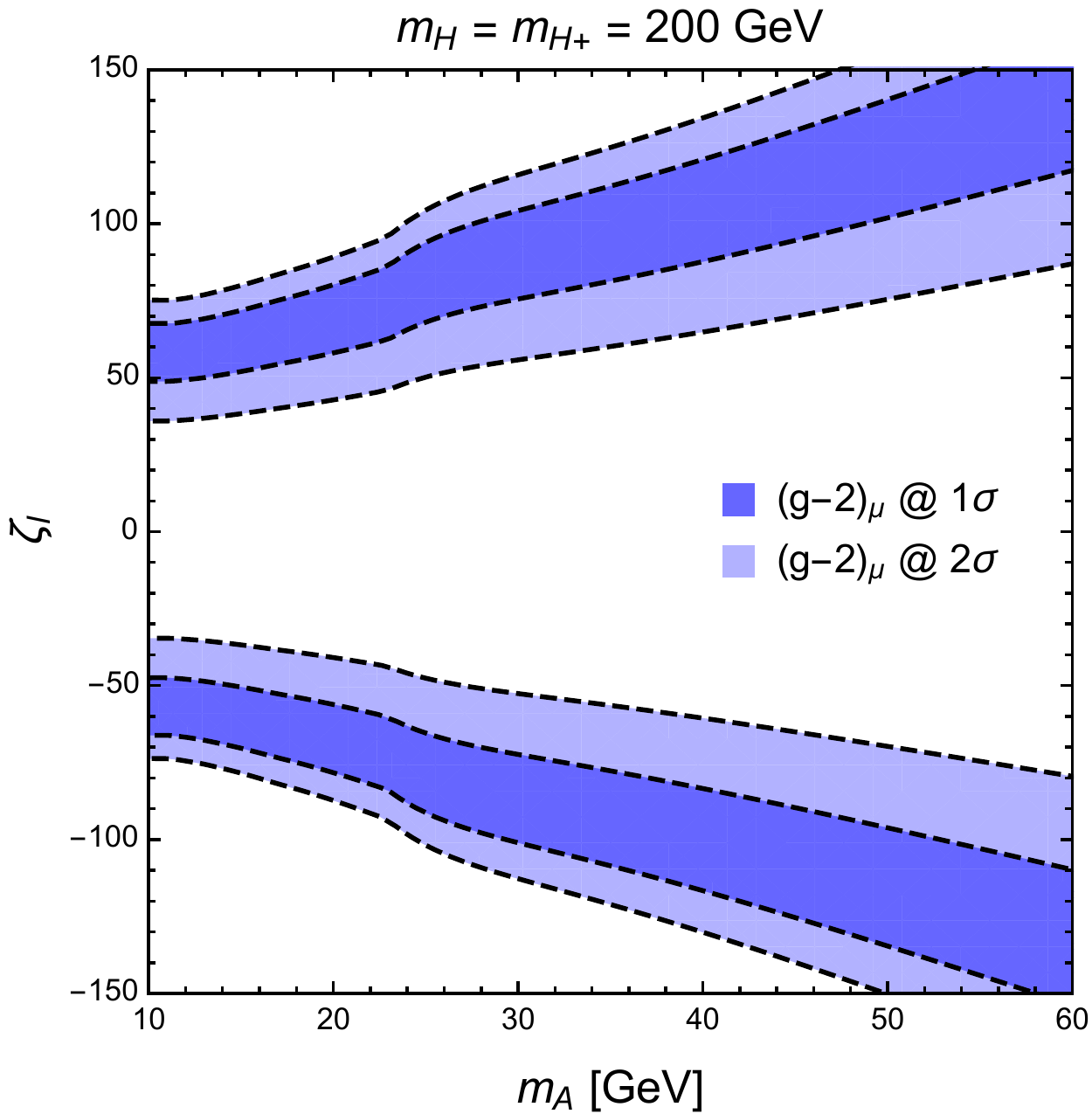}
\caption{The 1 and $2\sigma$ regions of the anomalous magnetic moment of the muon in the parameter space of $m_A$ and $\zeta_\ell$. For the sake of definiteness, the mass of the charged Higgs has been chosen as $m_{H^\pm} = 200$ GeV.}
\label{AMM-muon}
\end{figure}

\begin{figure}
\includegraphics[scale=0.45]{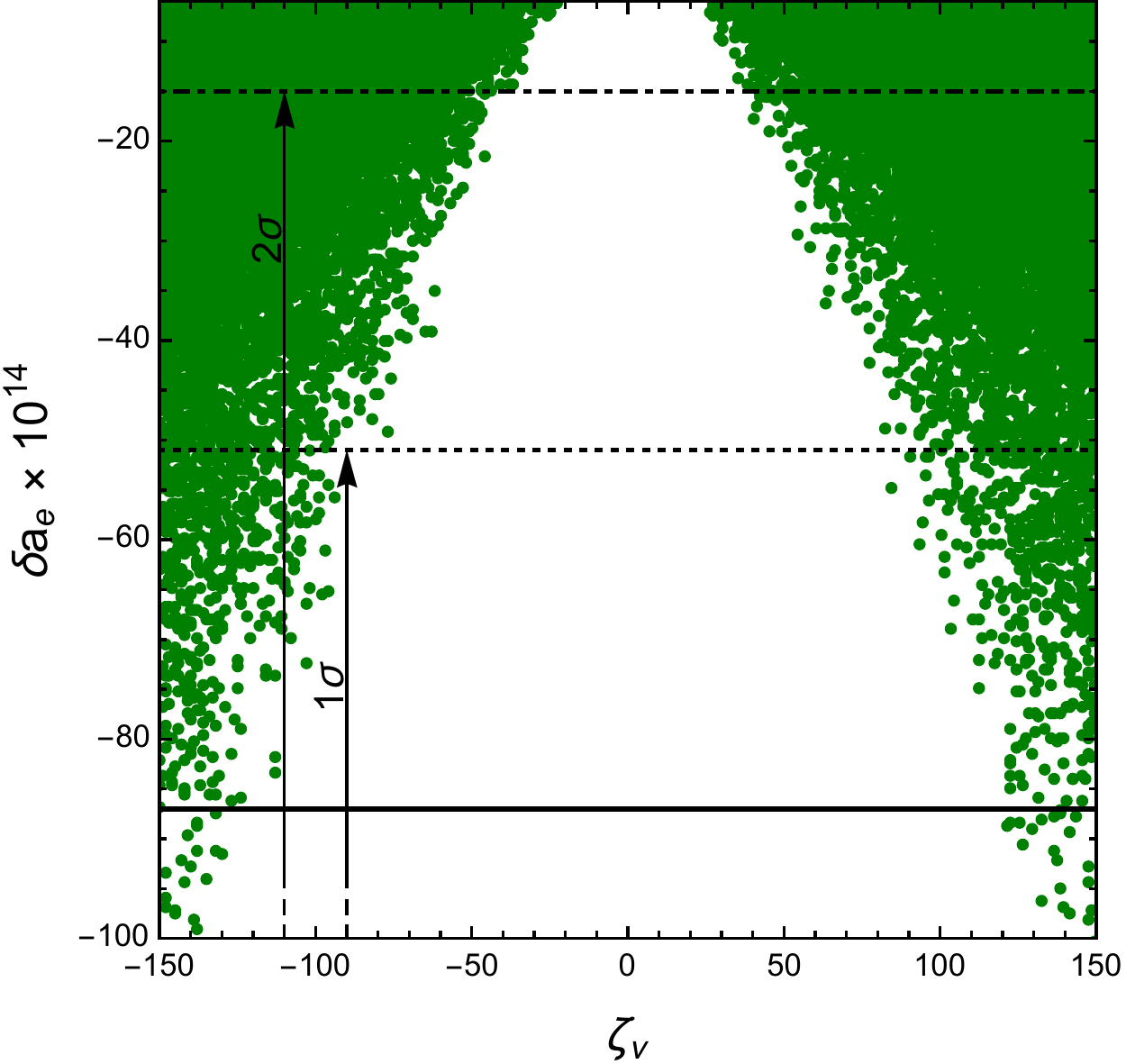}
\includegraphics[scale=0.45]{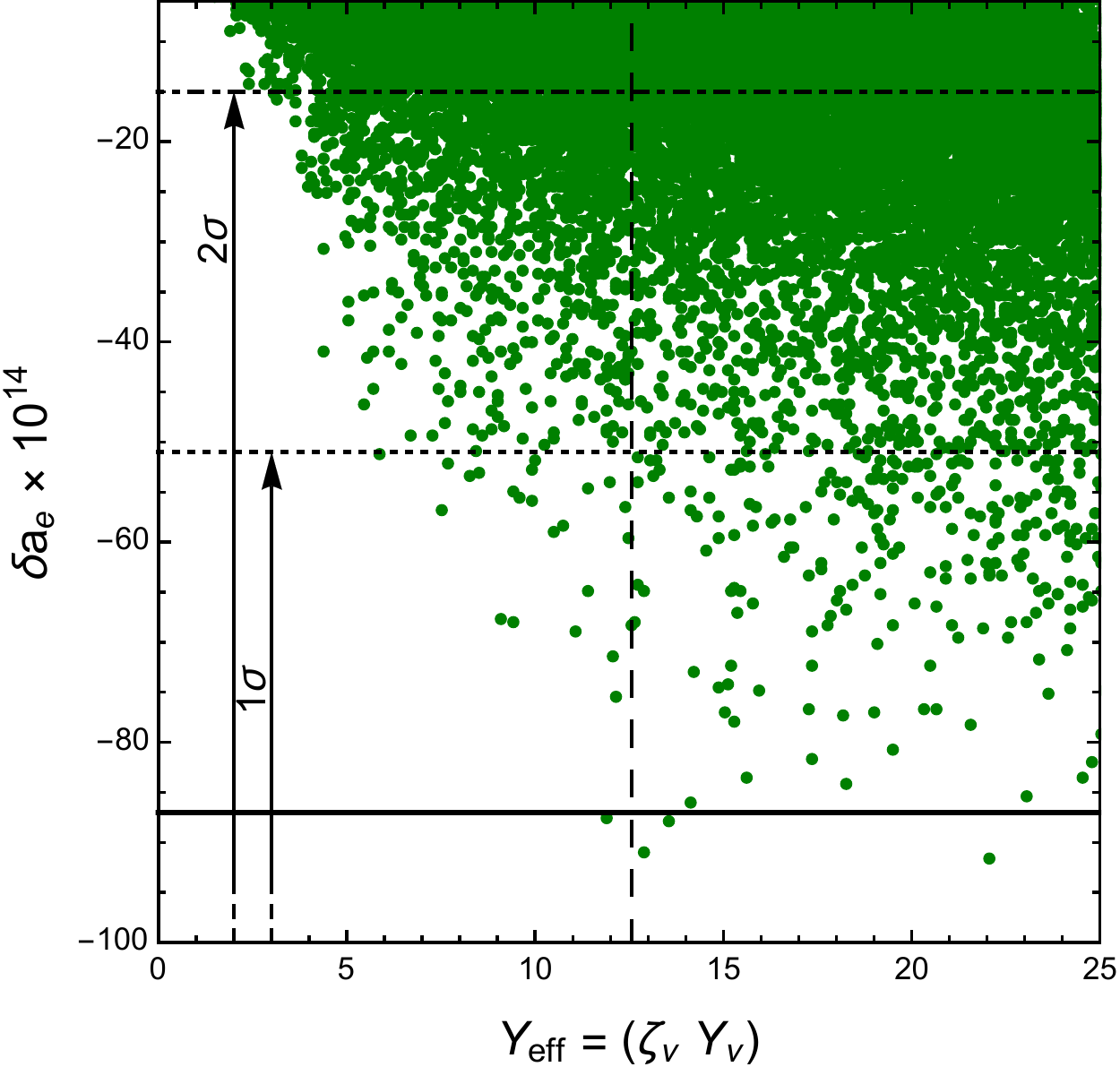}
\includegraphics[scale=0.45]{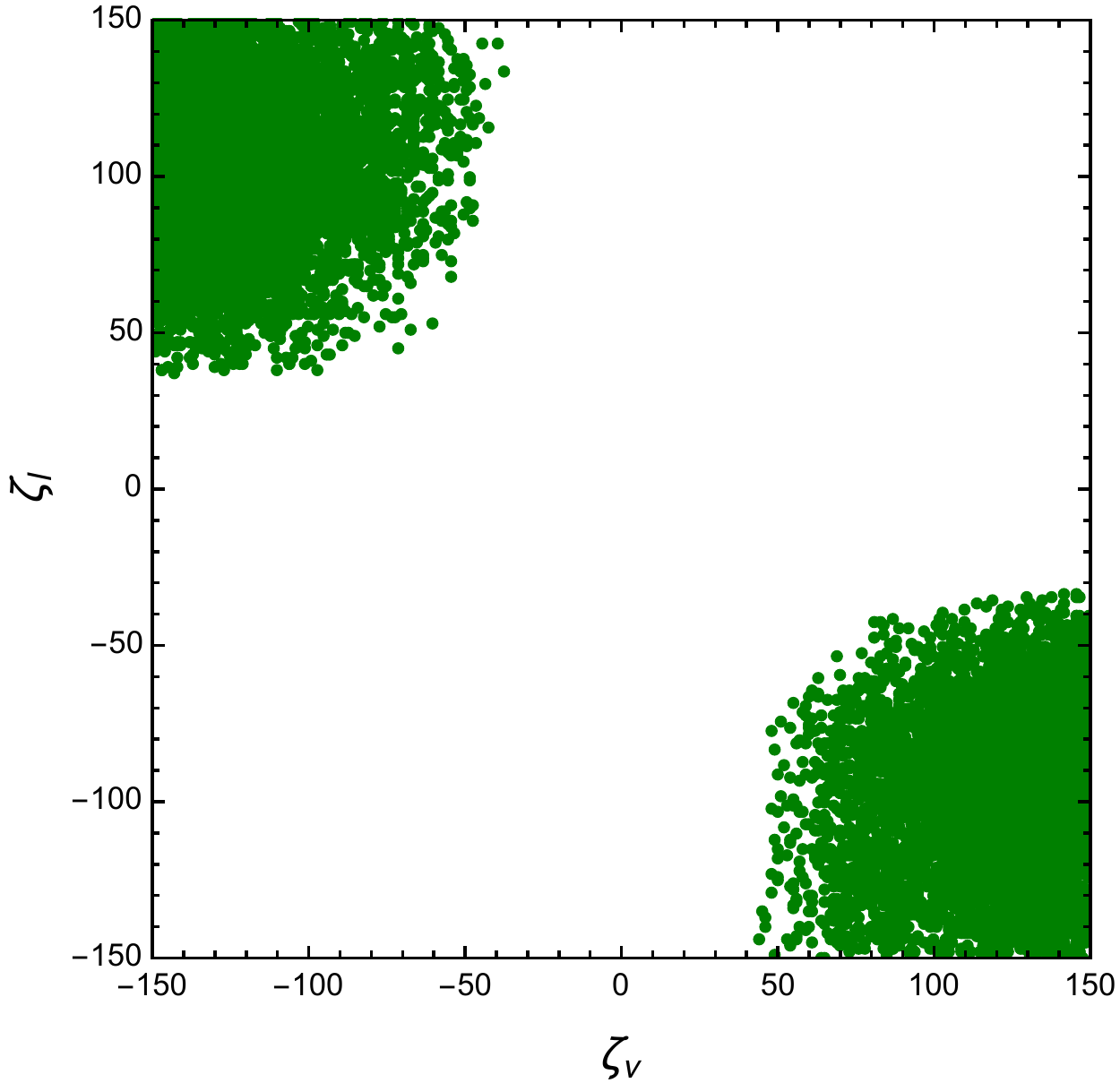}
\caption{The anomalous magnetic moment of the electron as a function of (a) $\zeta_\nu$ and (b) the effective neutrino coupling $\zeta_\nu Y_\nu$. The horizontal solid, dashed and dot-dashed lines correspond, respectively, to the central value, the upper $1\sigma$ band and the upper $2\sigma$ band. The vertical dashed line in  (b) represents the maximum allowed value for $Y_\textrm{eff}=\zeta_\nu Y_\nu$ from perturbativity. All the points satisfy the experimental and theoretical constraints as explained in the text and reproduce $a_\mu$ at $2\sigma$ level. (c) Distribution of points in the $(\zeta_\nu, \zeta_\ell)$ plane complying with all current experimental and theoretical bounds as well as with the solution of the $a_e$ and $a_\mu$ anomalies at $2\sigma$.}
\label{AMM-electron}
\end{figure}

The results of our analysis are depicted in Figs.~\ref{AMM-muon} and \ref{AMM-electron}. The former shows the regions in which the predicted $a_\mu$ is within 1 and $2\sigma$ around the measured central value. These are projected onto the most relevant parameter space defined by $m_A$ and $\zeta_\ell$. The mass of the charged Higgs boson has been fixed at a reference value of $m_{H^\pm} = 200$ GeV. Different choices of $m_{H^\pm}$ slightly modify the contours shown in the plot. In Fig.~\ref{AMM-electron} we show the prediction for $a_e$.
The points are generated by scanning over the parameter space of the model and comply with the experimental and theoretical bounds quoted above while  reproducing $a_\mu$ within the $2\sigma$ range. The parameters are scanned as follows: $m_{\nu_{h}} \in (200, 2000)$ GeV, $m_{H\pm}, m_H \in (100,1000)$ GeV, $m_A \in (10,60)$ GeV, $\zeta_\ell, \zeta_\nu \in (-150,150)$ and $|(U_{Lh})_{\mu \, \nu_h}|^2 \in (10^{-5}, 10^{-3})$. In Fig.~\ref{AMM-electron}(a) and (b),  $(g-2)_e$ is plotted, respectively, against $\zeta_\nu$ and the effective coupling $\zeta_\nu Y_\nu$ which characterises this model and that has been extensively discussed in \cite{DelleRose:2019ukt}. The vertical dashed line shows the maximum allowed value required by pertubativity. Finally, Fig.~\ref{AMM-electron}(c)  shows the distribution of points along the $\zeta_\nu$ and $\zeta_\ell$ directions compliant with all the bounds discussed above as well as $a_e$ and $a_\mu$ measurements within $2\sigma$. As mentioned already, the two couplings must necessarily have opposite signs.

%%%%%%%%%%%%%%%%%%%%%%%%%
\subsection{LHC phenomenology of the extra (pseudo)scalar bosons}

\begin{figure}
\includegraphics[scale=0.45]{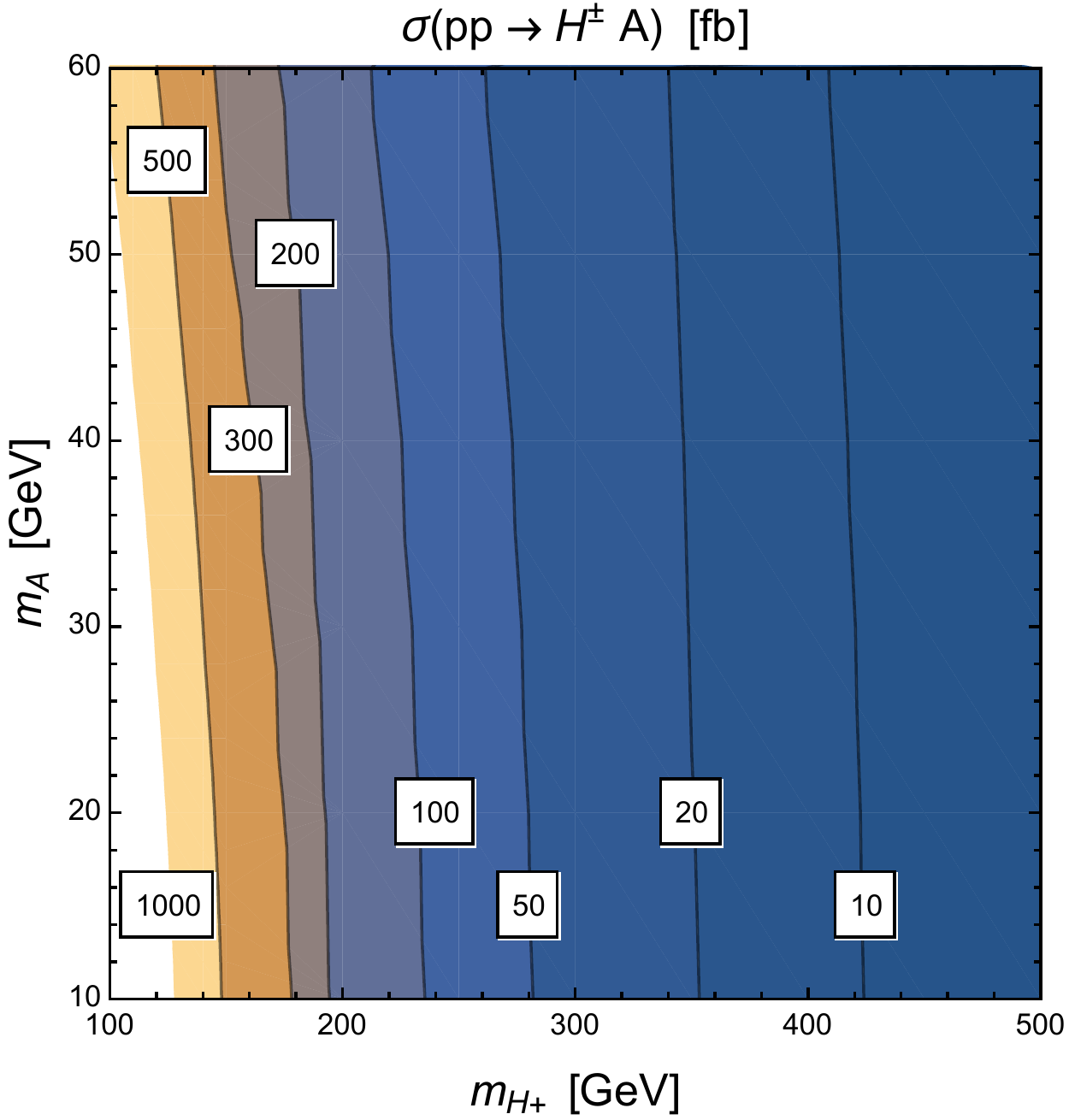}
\includegraphics[scale=0.45]{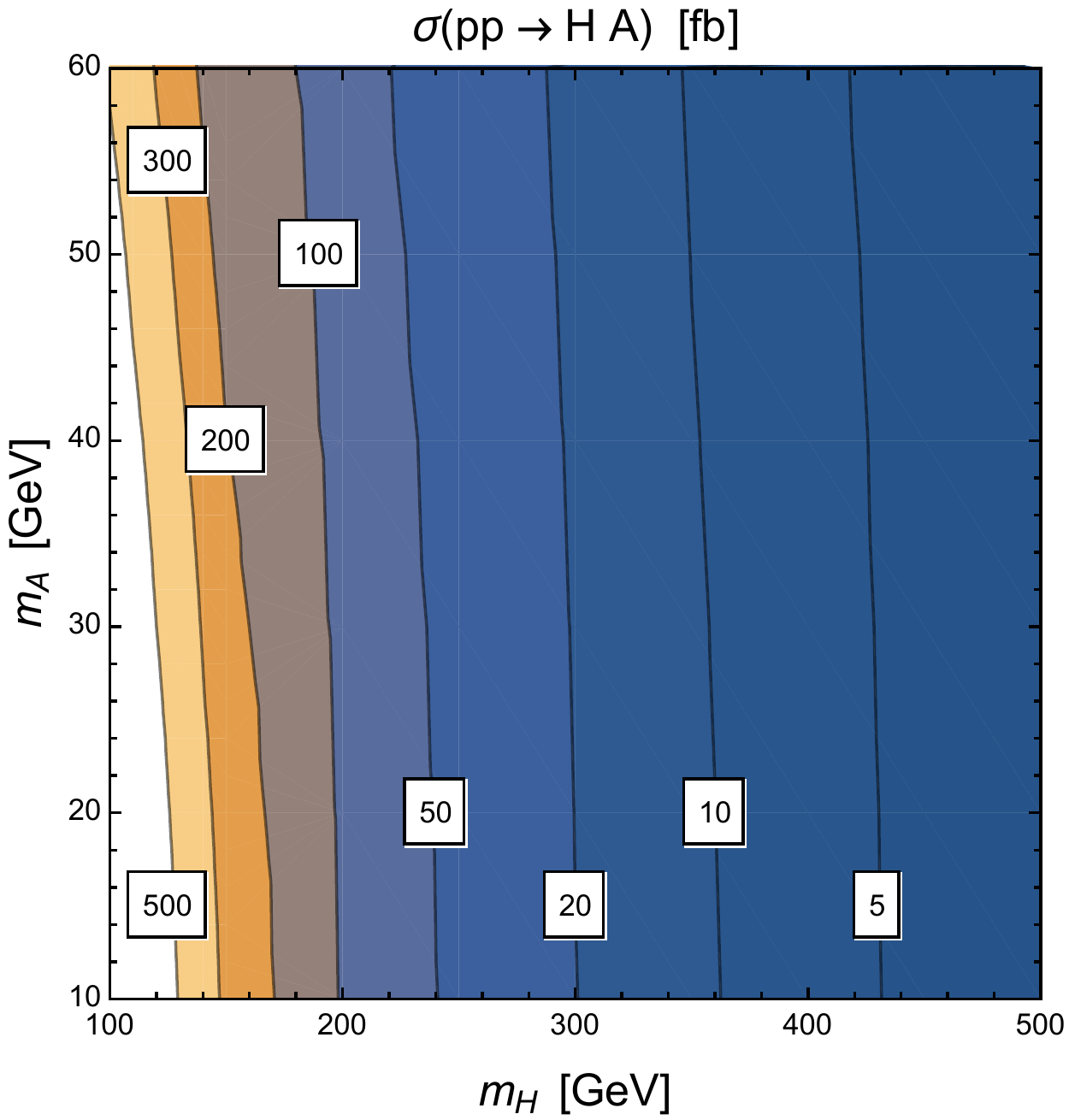}
\includegraphics[scale=0.45]{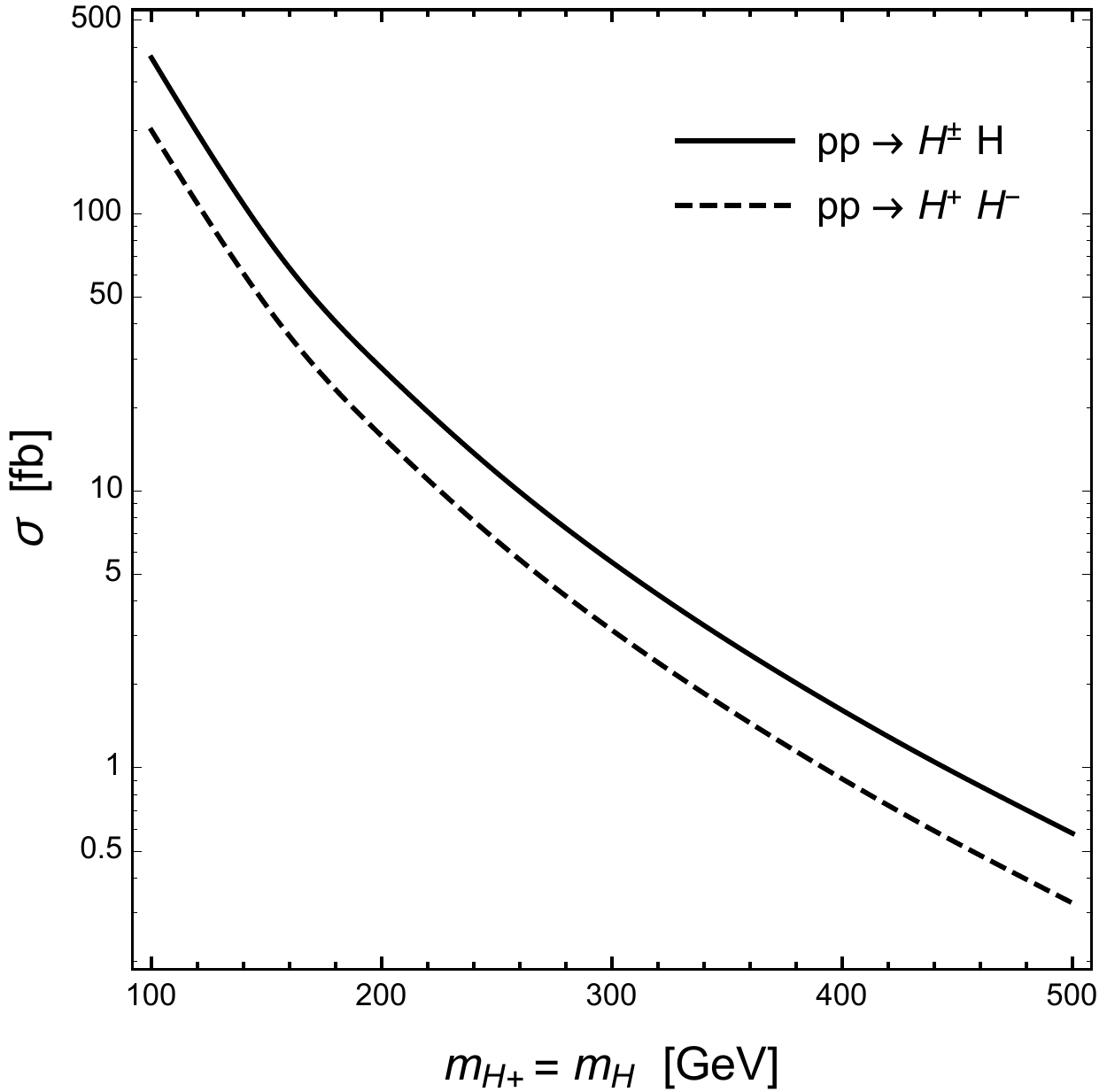}
\caption{The LHC production cross sections of pairs of the extra Higgs bosons as functions of $m_A$ and $m_{H^\pm} = m_H$. \label{fig:xs}}
\end{figure}

In the leptophilic scenario delineated above, the light pseudoscalar state $A$ can decay at tree-level via $A \to \tau \tau$ with BR close to  $100\%$. For the charged Higgs boson, instead, the two main open decay modes are $H^\pm \to A W^\pm$, where the interaction is completely fixed by the $SU(2)_L$ gauge coupling, and $H^\pm \to \tau^\pm \nu$, which is controlled by the $\zeta_\ell$ coupling. Analogously, for the heavy neutral scalar state $H$ the two leading decay modes are $H \to \tau \tau$ and $H \to A Z$.
For large $m_{H^\pm}, m_H$, the BRs of the $H^\pm$ and $H$ are solely controlled by the coupling $g_\ell = \zeta_\ell \, m_\tau / m_{H^\pm}$ and are approximated by\footnote{We neglected small deviations from $\sin(\beta-\alpha) = 1$.}
\bea
\textrm{BR}(H^\pm \to A W^\pm) = \textrm{BR}(H \to A Z) = \frac{1}{1 + 2 g_\ell^2} \,, \qquad  \textrm{BR}(H^\pm \to \tau^\pm \nu) = \textrm{BR}(H \to \tau \tau) = \frac{2 g_\ell^2}{1 + 2 g_\ell^2}   \,.
\eea
Since the couplings to the quarks are suppressed, the main production modes proceed through the EW interactions. The relevant processes are
\bea
pp \to H^\pm A \,, \qquad pp \to H A \,, \qquad  pp \to H^\pm H \,, \qquad  pp \to H^+ H^-,
\eea
with the corresponding cross sections being only functions of the masses of the corresponding particles. 
The cross sections at the LHC are computed with MadGraph  \cite{Alwall:2014hca} and are shown in Fig.~\ref{fig:xs}. The largest contributions arise from $H^\pm A$ and $H A$.

The main signatures resulting from  these processes are characterised by final states with several $\tau$ leptons
\bea
3 \tau + \slashed{E}_T, \qquad 4 \tau + W^\pm, \qquad 4 \tau  , \qquad 4 \tau + Z,
\eea
where the first two stem from  $H^\pm A$ production (with a subleading component from $H^\pm H$) while the last two arise from the $H A$ production. A thorough analysis is beyond the scope of this paper. In order to get a feeling of the potential of these channels, here we list only an estimate of the inclusive cross section for the corresponding SM background
\bea
& \sigma_\textrm{SM}(Z W^\pm \to 3 \tau + \slashed{E}_T) \simeq 94 \, \textrm{fb}, \qquad 
& \sigma_\textrm{SM}(Z Z W^\pm \to 4 \tau + W^\pm) \simeq 3.2 \times 10^{-2} \, \textrm{fb}, \nn \\ 
& \sigma_\textrm{SM}(Z Z \to 4 \tau) \simeq 11 \, \textrm{fb}, \qquad 
& \sigma_\textrm{SM}(Z Z Z \to 4 \tau + Z ) \simeq 1.1 \times 10^{-2} \, \textrm{fb} \,.
\eea

\section{Conclusions}

The measurements of the the anomalous magnetic moment of electron and muon are amongst the most precise ones in the whole of particle physics, probing not only the structure of the SM but also the possibility of BSM theories entering these experimental observables. Intriguingly, both of these are currently showing some anomalies with respect to the SM predictions. Crucially, the two results go in different directions, i.e., the measurement of $a_{\mu}$ exceeds the SM result while that of $a_e$ lies below the corresponding SM yield. This circumstance makes it difficult to  find BSM solutions, as multiple new particles are generally needed, each contributing its corrections in different directions, i.e., with different signs, unless significant violation of discrete quantum numbers is exploited. 

In this paper,  we adopted  an A2HDM supplemented by RH neutrinos, respecting all the SM symmetries. In such a BSM framework, a possible explanation to the aforementioned anomalies can be attained  through one and two-loop topologies wherein the contribution from a very light CP-odd neutral Higgs state interacting with leptons, is tensioned against the one due to a charged Higgs boson interacting with the new heavy neutrinos, the latter with mass at the EW scale. Crucially, such a spectrum is able to explain the two leptonic anomalous magnetic moment measurements while also predicting new hallmark signals in the form of  $q\bar q' \to H^\pm A$ production yielding multi-$\tau$ final states, which are almost background free at the LHC and thus accessible already with current data samples. 

%%%%%%%%%%%%%%%%%%%%%%%%%

\section*{Acknowledgements}
SM is financed in part through
the NExT Institute and the STFC Consolidated Grant No. ST/L000296/1. 
LDR acknowledges support by the Spanish Ministry MEC under grant FPA 2017-88915-P and the Severo Ochoa excellence program of MINECO (SEV-2016-0588). IFAE is partially funded by the CERCA program of the Generalitat de Catalunya. The project that gave rise to these results received the support of a fellowship from ''la Caixa'' Foundation (ID 100010434) and from the European Union's Horizon 2020 research and innovation programme under the Marie Sklodowska-Curie Action grant agreement No 847648. The fellowship code is LCF/BQ/PI20/11760032.

\bibliographystyle{apsrev4-1}
\bibliography{draftbib}

\begin{thebibliography}{47}%
\makeatletter
\providecommand \@ifxundefined [1]{%
 \@ifx{#1\undefined}
}%
\providecommand \@ifnum [1]{%
 \ifnum #1\expandafter \@firstoftwo
 \else \expandafter \@secondoftwo
 \fi
}%
\providecommand \@ifx [1]{%
 \ifx #1\expandafter \@firstoftwo
 \else \expandafter \@secondoftwo
 \fi
}%
\providecommand \natexlab [1]{#1}%
\providecommand \enquote  [1]{``#1''}%
\providecommand \bibnamefont  [1]{#1}%
\providecommand \bibfnamefont [1]{#1}%
\providecommand \citenamefont [1]{#1}%
\providecommand \href@noop [0]{\@secondoftwo}%
\providecommand \href [0]{\begingroup \@sanitize@url \@href}%
\providecommand \@href[1]{\@@startlink{#1}\@@href}%
\providecommand \@@href[1]{\endgroup#1\@@endlink}%
\providecommand \@sanitize@url [0]{\catcode `\\12\catcode `\$12\catcode
  `\&12\catcode `\#12\catcode `\^12\catcode `\_12\catcode `\%12\relax}%
\providecommand \@@startlink[1]{}%
\providecommand \@@endlink[0]{}%
\providecommand \url  [0]{\begingroup\@sanitize@url \@url }%
\providecommand \@url [1]{\endgroup\@href {#1}{\urlprefix }}%
\providecommand \urlprefix  [0]{URL }%
\providecommand \Eprint [0]{\href }%
\providecommand \doibase [0]{http://dx.doi.org/}%
\providecommand \selectlanguage [0]{\@gobble}%
\providecommand \bibinfo  [0]{\@secondoftwo}%
\providecommand \bibfield  [0]{\@secondoftwo}%
\providecommand \translation [1]{[#1]}%
\providecommand \BibitemOpen [0]{}%
\providecommand \bibitemStop [0]{}%
\providecommand \bibitemNoStop [0]{.\EOS\space}%
\providecommand \EOS [0]{\spacefactor3000\relax}%
\providecommand \BibitemShut  [1]{\csname bibitem#1\endcsname}%
\let\auto@bib@innerbib\@empty
%</preamble>
\bibitem [{\citenamefont {Keshavarzi}\ \emph {et~al.}(2020)\citenamefont
  {Keshavarzi}, \citenamefont {Marciano}, \citenamefont {Passera},\ and\
  \citenamefont {Sirlin}}]{Keshavarzi:2020bfy}%
  \BibitemOpen
  \bibfield  {author} {\bibinfo {author} {\bibfnamefont {A.}~\bibnamefont
  {Keshavarzi}}, \bibinfo {author} {\bibfnamefont {W.~J.}\ \bibnamefont
  {Marciano}}, \bibinfo {author} {\bibfnamefont {M.}~\bibnamefont {Passera}}, \
  and\ \bibinfo {author} {\bibfnamefont {A.}~\bibnamefont {Sirlin}},\ }\href
  {\doibase 10.1103/PhysRevD.102.033002} {\bibfield  {journal} {\bibinfo
  {journal} {Phys. Rev. D}\ }\textbf {\bibinfo {volume} {102}},\ \bibinfo
  {pages} {033002} (\bibinfo {year} {2020})},\ \Eprint
  {http://arxiv.org/abs/2006.12666} {arXiv:2006.12666 [hep-ph]} \BibitemShut
  {NoStop}%
\bibitem [{\citenamefont {Parker}\ \emph {et~al.}(2018)\citenamefont {Parker},
  \citenamefont {Yu}, \citenamefont {Zhong}, \citenamefont {Estey},\ and\
  \citenamefont {M\"uller}}]{Parker:2018vye}%
  \BibitemOpen
  \bibfield  {author} {\bibinfo {author} {\bibfnamefont {R.~H.}\ \bibnamefont
  {Parker}}, \bibinfo {author} {\bibfnamefont {C.}~\bibnamefont {Yu}}, \bibinfo
  {author} {\bibfnamefont {W.}~\bibnamefont {Zhong}}, \bibinfo {author}
  {\bibfnamefont {B.}~\bibnamefont {Estey}}, \ and\ \bibinfo {author}
  {\bibfnamefont {H.}~\bibnamefont {M\"uller}},\ }\href {\doibase
  10.1126/science.aap7706} {\bibfield  {journal} {\bibinfo  {journal}
  {Science}\ }\textbf {\bibinfo {volume} {360}},\ \bibinfo {pages} {191}
  (\bibinfo {year} {2018})},\ \Eprint {http://arxiv.org/abs/1812.04130}
  {arXiv:1812.04130 [physics.atom-ph]} \BibitemShut {NoStop}%
\bibitem [{\citenamefont {Semertzidis}\ \emph {et~al.}(1999)\citenamefont
  {Semertzidis} \emph {et~al.}}]{Semertzidis:1999kv}%
  \BibitemOpen
  \bibfield  {author} {\bibinfo {author} {\bibfnamefont {Y.}~\bibnamefont
  {Semertzidis}} \emph {et~al.},\ }in\ \href {\doibase
  10.1142/9789812791849_0007} {\emph {\bibinfo {booktitle} {{KEK International
  Workshop on High Intensity Muon Sources (HIMUS 99)}}}}\ (\bibinfo {year}
  {1999})\ pp.\ \bibinfo {pages} {81--96},\ \Eprint
  {http://arxiv.org/abs/hep-ph/0012087} {arXiv:hep-ph/0012087} \BibitemShut
  {NoStop}%
\bibitem [{\citenamefont {Farley}\ \emph {et~al.}(2004)\citenamefont {Farley},
  \citenamefont {Jungmann}, \citenamefont {Miller}, \citenamefont {Morse},
  \citenamefont {Orlov}, \citenamefont {Roberts}, \citenamefont {Semertzidis},
  \citenamefont {Silenko},\ and\ \citenamefont {Stephenson}}]{Farley:2003wt}%
  \BibitemOpen
  \bibfield  {author} {\bibinfo {author} {\bibfnamefont {F.}~\bibnamefont
  {Farley}}, \bibinfo {author} {\bibfnamefont {K.}~\bibnamefont {Jungmann}},
  \bibinfo {author} {\bibfnamefont {J.}~\bibnamefont {Miller}}, \bibinfo
  {author} {\bibfnamefont {W.}~\bibnamefont {Morse}}, \bibinfo {author}
  {\bibfnamefont {Y.}~\bibnamefont {Orlov}}, \bibinfo {author} {\bibfnamefont
  {B.}~\bibnamefont {Roberts}}, \bibinfo {author} {\bibfnamefont
  {Y.}~\bibnamefont {Semertzidis}}, \bibinfo {author} {\bibfnamefont
  {A.}~\bibnamefont {Silenko}}, \ and\ \bibinfo {author} {\bibfnamefont
  {E.}~\bibnamefont {Stephenson}},\ }\href {\doibase
  10.1103/PhysRevLett.93.052001} {\bibfield  {journal} {\bibinfo  {journal}
  {Phys. Rev. Lett.}\ }\textbf {\bibinfo {volume} {93}},\ \bibinfo {pages}
  {052001} (\bibinfo {year} {2004})},\ \Eprint
  {http://arxiv.org/abs/hep-ex/0307006} {arXiv:hep-ex/0307006} \BibitemShut
  {NoStop}%
\bibitem [{\citenamefont {Liu}\ \emph {et~al.}(2019)\citenamefont {Liu},
  \citenamefont {Wagner},\ and\ \citenamefont {Wang}}]{Liu:2018xkx}%
  \BibitemOpen
  \bibfield  {author} {\bibinfo {author} {\bibfnamefont {J.}~\bibnamefont
  {Liu}}, \bibinfo {author} {\bibfnamefont {C.~E.}\ \bibnamefont {Wagner}}, \
  and\ \bibinfo {author} {\bibfnamefont {X.-P.}\ \bibnamefont {Wang}},\ }\href
  {\doibase 10.1007/JHEP03(2019)008} {\bibfield  {journal} {\bibinfo  {journal}
  {JHEP}\ }\textbf {\bibinfo {volume} {03}},\ \bibinfo {pages} {008} (\bibinfo
  {year} {2019})},\ \Eprint {http://arxiv.org/abs/1810.11028} {arXiv:1810.11028
  [hep-ph]} \BibitemShut {NoStop}%
\bibitem [{\citenamefont {Han}\ \emph {et~al.}(2019)\citenamefont {Han},
  \citenamefont {Li}, \citenamefont {Wang},\ and\ \citenamefont
  {Zhang}}]{Han:2018znu}%
  \BibitemOpen
  \bibfield  {author} {\bibinfo {author} {\bibfnamefont {X.-F.}\ \bibnamefont
  {Han}}, \bibinfo {author} {\bibfnamefont {T.}~\bibnamefont {Li}}, \bibinfo
  {author} {\bibfnamefont {L.}~\bibnamefont {Wang}}, \ and\ \bibinfo {author}
  {\bibfnamefont {Y.}~\bibnamefont {Zhang}},\ }\href {\doibase
  10.1103/PhysRevD.99.095034} {\bibfield  {journal} {\bibinfo  {journal} {Phys.
  Rev. D}\ }\textbf {\bibinfo {volume} {99}},\ \bibinfo {pages} {095034}
  (\bibinfo {year} {2019})},\ \Eprint {http://arxiv.org/abs/1812.02449}
  {arXiv:1812.02449 [hep-ph]} \BibitemShut {NoStop}%
\bibitem [{\citenamefont {Endo}\ and\ \citenamefont
  {Yin}(2019)}]{Endo:2019bcj}%
  \BibitemOpen
  \bibfield  {author} {\bibinfo {author} {\bibfnamefont {M.}~\bibnamefont
  {Endo}}\ and\ \bibinfo {author} {\bibfnamefont {W.}~\bibnamefont {Yin}},\
  }\href {\doibase 10.1007/JHEP08(2019)122} {\bibfield  {journal} {\bibinfo
  {journal} {JHEP}\ }\textbf {\bibinfo {volume} {08}},\ \bibinfo {pages} {122}
  (\bibinfo {year} {2019})},\ \Eprint {http://arxiv.org/abs/1906.08768}
  {arXiv:1906.08768 [hep-ph]} \BibitemShut {NoStop}%
\bibitem [{\citenamefont {Bauer}\ \emph {et~al.}(2020)\citenamefont {Bauer},
  \citenamefont {Neubert}, \citenamefont {Renner}, \citenamefont {Schnubel},\
  and\ \citenamefont {Thamm}}]{Bauer:2019gfk}%
  \BibitemOpen
  \bibfield  {author} {\bibinfo {author} {\bibfnamefont {M.}~\bibnamefont
  {Bauer}}, \bibinfo {author} {\bibfnamefont {M.}~\bibnamefont {Neubert}},
  \bibinfo {author} {\bibfnamefont {S.}~\bibnamefont {Renner}}, \bibinfo
  {author} {\bibfnamefont {M.}~\bibnamefont {Schnubel}}, \ and\ \bibinfo
  {author} {\bibfnamefont {A.}~\bibnamefont {Thamm}},\ }\href {\doibase
  10.1103/PhysRevLett.124.211803} {\bibfield  {journal} {\bibinfo  {journal}
  {Phys. Rev. Lett.}\ }\textbf {\bibinfo {volume} {124}},\ \bibinfo {pages}
  {211803} (\bibinfo {year} {2020})},\ \Eprint
  {http://arxiv.org/abs/1908.00008} {arXiv:1908.00008 [hep-ph]} \BibitemShut
  {NoStop}%
\bibitem [{\citenamefont {Badziak}\ and\ \citenamefont
  {Sakurai}(2019)}]{Badziak:2019gaf}%
  \BibitemOpen
  \bibfield  {author} {\bibinfo {author} {\bibfnamefont {M.}~\bibnamefont
  {Badziak}}\ and\ \bibinfo {author} {\bibfnamefont {K.}~\bibnamefont
  {Sakurai}},\ }\href {\doibase 10.1007/JHEP10(2019)024} {\bibfield  {journal}
  {\bibinfo  {journal} {JHEP}\ }\textbf {\bibinfo {volume} {10}},\ \bibinfo
  {pages} {024} (\bibinfo {year} {2019})},\ \Eprint
  {http://arxiv.org/abs/1908.03607} {arXiv:1908.03607 [hep-ph]} \BibitemShut
  {NoStop}%
\bibitem [{\citenamefont {C\'arcamo~Hern\'andez}\ \emph
  {et~al.}(2020{\natexlab{a}})\citenamefont {C\'arcamo~Hern\'andez},
  \citenamefont {Hidalgo~Vel\'asquez}, \citenamefont {Kovalenko}, \citenamefont
  {Long}, \citenamefont {P\'erez-Julve},\ and\ \citenamefont
  {Vien}}]{CarcamoHernandez:2020pxw}%
  \BibitemOpen
  \bibfield  {author} {\bibinfo {author} {\bibfnamefont {A.}~\bibnamefont
  {C\'arcamo~Hern\'andez}}, \bibinfo {author} {\bibfnamefont {Y.}~\bibnamefont
  {Hidalgo~Vel\'asquez}}, \bibinfo {author} {\bibfnamefont {S.}~\bibnamefont
  {Kovalenko}}, \bibinfo {author} {\bibfnamefont {H.}~\bibnamefont {Long}},
  \bibinfo {author} {\bibfnamefont {N.~A.}\ \bibnamefont {P\'erez-Julve}}, \
  and\ \bibinfo {author} {\bibfnamefont {V.}~\bibnamefont {Vien}},\ }\href@noop
  {} {\  (\bibinfo {year} {2020}{\natexlab{a}})},\ \Eprint
  {http://arxiv.org/abs/2002.07347} {arXiv:2002.07347 [hep-ph]} \BibitemShut
  {NoStop}%
\bibitem [{\citenamefont {Haba}\ \emph {et~al.}(2020)\citenamefont {Haba},
  \citenamefont {Shimizu},\ and\ \citenamefont {Yamada}}]{Haba:2020gkr}%
  \BibitemOpen
  \bibfield  {author} {\bibinfo {author} {\bibfnamefont {N.}~\bibnamefont
  {Haba}}, \bibinfo {author} {\bibfnamefont {Y.}~\bibnamefont {Shimizu}}, \
  and\ \bibinfo {author} {\bibfnamefont {T.}~\bibnamefont {Yamada}},\ }\href
  {\doibase 10.1093/ptep/ptaa098} {\bibfield  {journal} {\bibinfo  {journal}
  {PTEP}\ }\textbf {\bibinfo {volume} {2020}},\ \bibinfo {pages} {093B05}
  (\bibinfo {year} {2020})},\ \Eprint {http://arxiv.org/abs/2002.10230}
  {arXiv:2002.10230 [hep-ph]} \BibitemShut {NoStop}%
\bibitem [{\citenamefont {Bigaran}\ and\ \citenamefont
  {Volkas}(2020)}]{Bigaran:2020jil}%
  \BibitemOpen
  \bibfield  {author} {\bibinfo {author} {\bibfnamefont {I.}~\bibnamefont
  {Bigaran}}\ and\ \bibinfo {author} {\bibfnamefont {R.~R.}\ \bibnamefont
  {Volkas}},\ }\href {\doibase 10.1103/PhysRevD.102.075037} {\bibfield
  {journal} {\bibinfo  {journal} {Phys. Rev. D}\ }\textbf {\bibinfo {volume}
  {102}},\ \bibinfo {pages} {075037} (\bibinfo {year} {2020})},\ \Eprint
  {http://arxiv.org/abs/2002.12544} {arXiv:2002.12544 [hep-ph]} \BibitemShut
  {NoStop}%
\bibitem [{\citenamefont {Calibbi}\ \emph {et~al.}(2020)\citenamefont
  {Calibbi}, \citenamefont {L\'opez-Ib\'a\~nez}, \citenamefont {Melis},\ and\
  \citenamefont {Vives}}]{Calibbi:2020emz}%
  \BibitemOpen
  \bibfield  {author} {\bibinfo {author} {\bibfnamefont {L.}~\bibnamefont
  {Calibbi}}, \bibinfo {author} {\bibfnamefont {M.}~\bibnamefont
  {L\'opez-Ib\'a\~nez}}, \bibinfo {author} {\bibfnamefont {A.}~\bibnamefont
  {Melis}}, \ and\ \bibinfo {author} {\bibfnamefont {O.}~\bibnamefont
  {Vives}},\ }\href {\doibase 10.1007/JHEP06(2020)087} {\bibfield  {journal}
  {\bibinfo  {journal} {JHEP}\ }\textbf {\bibinfo {volume} {06}},\ \bibinfo
  {pages} {087} (\bibinfo {year} {2020})},\ \Eprint
  {http://arxiv.org/abs/2003.06633} {arXiv:2003.06633 [hep-ph]} \BibitemShut
  {NoStop}%
\bibitem [{\citenamefont {Chen}\ and\ \citenamefont
  {Nomura}(2020)}]{Chen:2020jvl}%
  \BibitemOpen
  \bibfield  {author} {\bibinfo {author} {\bibfnamefont {C.-H.}\ \bibnamefont
  {Chen}}\ and\ \bibinfo {author} {\bibfnamefont {T.}~\bibnamefont {Nomura}},\
  }\href@noop {} {\  (\bibinfo {year} {2020})},\ \Eprint
  {http://arxiv.org/abs/2003.07638} {arXiv:2003.07638 [hep-ph]} \BibitemShut
  {NoStop}%
\bibitem [{\citenamefont {Jana}\ \emph
  {et~al.}(2020{\natexlab{a}})\citenamefont {Jana}, \citenamefont {K.},\ and\
  \citenamefont {Saad}}]{Jana:2020pxx}%
  \BibitemOpen
  \bibfield  {author} {\bibinfo {author} {\bibfnamefont {S.}~\bibnamefont
  {Jana}}, \bibinfo {author} {\bibfnamefont {V.~P.}\ \bibnamefont {K.}}, \ and\
  \bibinfo {author} {\bibfnamefont {S.}~\bibnamefont {Saad}},\ }\href {\doibase
  10.1103/PhysRevD.101.115037} {\bibfield  {journal} {\bibinfo  {journal}
  {Phys. Rev. D}\ }\textbf {\bibinfo {volume} {101}},\ \bibinfo {pages}
  {115037} (\bibinfo {year} {2020}{\natexlab{a}})},\ \Eprint
  {http://arxiv.org/abs/2003.03386} {arXiv:2003.03386 [hep-ph]} \BibitemShut
  {NoStop}%
\bibitem [{\citenamefont {Li}\ \emph {et~al.}(2020)\citenamefont {Li},
  \citenamefont {Li}, \citenamefont {Li}, \citenamefont {Yang},\ and\
  \citenamefont {Zhang}}]{Li:2020dbg}%
  \BibitemOpen
  \bibfield  {author} {\bibinfo {author} {\bibfnamefont {S.-P.}\ \bibnamefont
  {Li}}, \bibinfo {author} {\bibfnamefont {X.-Q.}\ \bibnamefont {Li}}, \bibinfo
  {author} {\bibfnamefont {Y.-Y.}\ \bibnamefont {Li}}, \bibinfo {author}
  {\bibfnamefont {Y.-D.}\ \bibnamefont {Yang}}, \ and\ \bibinfo {author}
  {\bibfnamefont {X.}~\bibnamefont {Zhang}},\ }\href@noop {} {\  (\bibinfo
  {year} {2020})},\ \Eprint {http://arxiv.org/abs/2010.02799} {arXiv:2010.02799
  [hep-ph]} \BibitemShut {NoStop}%
\bibitem [{\citenamefont {Chun}\ and\ \citenamefont
  {Mondal}(2020)}]{Chun:2020uzw}%
  \BibitemOpen
  \bibfield  {author} {\bibinfo {author} {\bibfnamefont {E.~J.}\ \bibnamefont
  {Chun}}\ and\ \bibinfo {author} {\bibfnamefont {T.}~\bibnamefont {Mondal}},\
  }\href {\doibase 10.1007/JHEP11(2020)077} {\bibfield  {journal} {\bibinfo
  {journal} {JHEP}\ }\textbf {\bibinfo {volume} {11}},\ \bibinfo {pages} {077}
  (\bibinfo {year} {2020})},\ \Eprint {http://arxiv.org/abs/2009.08314}
  {arXiv:2009.08314 [hep-ph]} \BibitemShut {NoStop}%
\bibitem [{\citenamefont {Jana}\ \emph
  {et~al.}(2020{\natexlab{b}})\citenamefont {Jana}, \citenamefont {Vishnu},
  \citenamefont {Rodejohann},\ and\ \citenamefont {Saad}}]{Jana:2020joi}%
  \BibitemOpen
  \bibfield  {author} {\bibinfo {author} {\bibfnamefont {S.}~\bibnamefont
  {Jana}}, \bibinfo {author} {\bibfnamefont {P.}~\bibnamefont {Vishnu}},
  \bibinfo {author} {\bibfnamefont {W.}~\bibnamefont {Rodejohann}}, \ and\
  \bibinfo {author} {\bibfnamefont {S.}~\bibnamefont {Saad}},\ }\href {\doibase
  10.1103/PhysRevD.102.075003} {\bibfield  {journal} {\bibinfo  {journal}
  {Phys. Rev. D}\ }\textbf {\bibinfo {volume} {102}},\ \bibinfo {pages}
  {075003} (\bibinfo {year} {2020}{\natexlab{b}})},\ \Eprint
  {http://arxiv.org/abs/2008.02377} {arXiv:2008.02377 [hep-ph]} \BibitemShut
  {NoStop}%
\bibitem [{\citenamefont {Arbel\'aez}\ \emph {et~al.}(2020)\citenamefont
  {Arbel\'aez}, \citenamefont {Cepedello}, \citenamefont {Fonseca},\ and\
  \citenamefont {Hirsch}}]{Arbelaez:2020rbq}%
  \BibitemOpen
  \bibfield  {author} {\bibinfo {author} {\bibfnamefont {C.}~\bibnamefont
  {Arbel\'aez}}, \bibinfo {author} {\bibfnamefont {R.}~\bibnamefont
  {Cepedello}}, \bibinfo {author} {\bibfnamefont {R.~M.}\ \bibnamefont
  {Fonseca}}, \ and\ \bibinfo {author} {\bibfnamefont {M.}~\bibnamefont
  {Hirsch}},\ }\href {\doibase 10.1103/PhysRevD.102.075005} {\bibfield
  {journal} {\bibinfo  {journal} {Phys. Rev. D}\ }\textbf {\bibinfo {volume}
  {102}},\ \bibinfo {pages} {075005} (\bibinfo {year} {2020})},\ \Eprint
  {http://arxiv.org/abs/2007.11007} {arXiv:2007.11007 [hep-ph]} \BibitemShut
  {NoStop}%
\bibitem [{\citenamefont {Delle~Rose}\ \emph
  {et~al.}(2020{\natexlab{a}})\citenamefont {Delle~Rose}, \citenamefont
  {Marzo},\ and\ \citenamefont {Marzola}}]{DelleRose:2020qak}%
  \BibitemOpen
  \bibfield  {author} {\bibinfo {author} {\bibfnamefont {L.}~\bibnamefont
  {Delle~Rose}}, \bibinfo {author} {\bibfnamefont {C.}~\bibnamefont {Marzo}}, \
  and\ \bibinfo {author} {\bibfnamefont {L.}~\bibnamefont {Marzola}},\
  }\href@noop {} {\  (\bibinfo {year} {2020}{\natexlab{a}})},\ \Eprint
  {http://arxiv.org/abs/2005.12389} {arXiv:2005.12389 [hep-ph]} \BibitemShut
  {NoStop}%
\bibitem [{\citenamefont {Crivellin}\ \emph {et~al.}(2018)\citenamefont
  {Crivellin}, \citenamefont {Hoferichter},\ and\ \citenamefont
  {Schmidt-Wellenburg}}]{Crivellin:2018qmi}%
  \BibitemOpen
  \bibfield  {author} {\bibinfo {author} {\bibfnamefont {A.}~\bibnamefont
  {Crivellin}}, \bibinfo {author} {\bibfnamefont {M.}~\bibnamefont
  {Hoferichter}}, \ and\ \bibinfo {author} {\bibfnamefont {P.}~\bibnamefont
  {Schmidt-Wellenburg}},\ }\href {\doibase 10.1103/PhysRevD.98.113002}
  {\bibfield  {journal} {\bibinfo  {journal} {Phys. Rev. D}\ }\textbf {\bibinfo
  {volume} {98}},\ \bibinfo {pages} {113002} (\bibinfo {year} {2018})},\
  \Eprint {http://arxiv.org/abs/1807.11484} {arXiv:1807.11484 [hep-ph]}
  \BibitemShut {NoStop}%
\bibitem [{\citenamefont {Dutta}\ \emph {et~al.}(2020)\citenamefont {Dutta},
  \citenamefont {Ghosh},\ and\ \citenamefont {Li}}]{Dutta:2020scq}%
  \BibitemOpen
  \bibfield  {author} {\bibinfo {author} {\bibfnamefont {B.}~\bibnamefont
  {Dutta}}, \bibinfo {author} {\bibfnamefont {S.}~\bibnamefont {Ghosh}}, \ and\
  \bibinfo {author} {\bibfnamefont {T.}~\bibnamefont {Li}},\ }\href {\doibase
  10.1103/PhysRevD.102.055017} {\bibfield  {journal} {\bibinfo  {journal}
  {Phys. Rev. D}\ }\textbf {\bibinfo {volume} {102}},\ \bibinfo {pages}
  {055017} (\bibinfo {year} {2020})},\ \Eprint
  {http://arxiv.org/abs/2006.01319} {arXiv:2006.01319 [hep-ph]} \BibitemShut
  {NoStop}%
\bibitem [{\citenamefont {Hati}\ \emph {et~al.}(2020)\citenamefont {Hati},
  \citenamefont {Kriewald}, \citenamefont {Orloff},\ and\ \citenamefont
  {Teixeira}}]{Hati:2020fzp}%
  \BibitemOpen
  \bibfield  {author} {\bibinfo {author} {\bibfnamefont {C.}~\bibnamefont
  {Hati}}, \bibinfo {author} {\bibfnamefont {J.}~\bibnamefont {Kriewald}},
  \bibinfo {author} {\bibfnamefont {J.}~\bibnamefont {Orloff}}, \ and\ \bibinfo
  {author} {\bibfnamefont {A.}~\bibnamefont {Teixeira}},\ }\href {\doibase
  10.1007/JHEP07(2020)235} {\bibfield  {journal} {\bibinfo  {journal} {JHEP}\
  }\textbf {\bibinfo {volume} {07}},\ \bibinfo {pages} {235} (\bibinfo {year}
  {2020})},\ \Eprint {http://arxiv.org/abs/2005.00028} {arXiv:2005.00028
  [hep-ph]} \BibitemShut {NoStop}%
\bibitem [{\citenamefont {C\'arcamo~Hern\'andez}\ \emph
  {et~al.}(2020{\natexlab{b}})\citenamefont {C\'arcamo~Hern\'andez},
  \citenamefont {King}, \citenamefont {Lee},\ and\ \citenamefont
  {Rowley}}]{CarcamoHernandez:2019ydc}%
  \BibitemOpen
  \bibfield  {author} {\bibinfo {author} {\bibfnamefont {A.}~\bibnamefont
  {C\'arcamo~Hern\'andez}}, \bibinfo {author} {\bibfnamefont {S.}~\bibnamefont
  {King}}, \bibinfo {author} {\bibfnamefont {H.}~\bibnamefont {Lee}}, \ and\
  \bibinfo {author} {\bibfnamefont {S.}~\bibnamefont {Rowley}},\ }\href
  {\doibase 10.1103/PhysRevD.101.115016} {\bibfield  {journal} {\bibinfo
  {journal} {Phys. Rev. D}\ }\textbf {\bibinfo {volume} {101}},\ \bibinfo
  {pages} {115016} (\bibinfo {year} {2020}{\natexlab{b}})},\ \Eprint
  {http://arxiv.org/abs/1910.10734} {arXiv:1910.10734 [hep-ph]} \BibitemShut
  {NoStop}%
\bibitem [{\citenamefont {Crivellin}\ and\ \citenamefont
  {Hoferichter}(2019)}]{Crivellin:2019mvj}%
  \BibitemOpen
  \bibfield  {author} {\bibinfo {author} {\bibfnamefont {A.}~\bibnamefont
  {Crivellin}}\ and\ \bibinfo {author} {\bibfnamefont {M.}~\bibnamefont
  {Hoferichter}},\ }in\ \href@noop {} {\emph {\bibinfo {booktitle} {{An Alpine
  LHC Physics Summit 2019}}}}\ (\bibinfo {year} {2019})\ pp.\ \bibinfo {pages}
  {29--34},\ \Eprint {http://arxiv.org/abs/1905.03789} {arXiv:1905.03789
  [hep-ph]} \BibitemShut {NoStop}%
\bibitem [{\citenamefont {Botella}\ \emph {et~al.}(2020)\citenamefont
  {Botella}, \citenamefont {Cornet-Gomez},\ and\ \citenamefont
  {Nebot}}]{Botella:2020xzf}%
  \BibitemOpen
  \bibfield  {author} {\bibinfo {author} {\bibfnamefont {F.~J.}\ \bibnamefont
  {Botella}}, \bibinfo {author} {\bibfnamefont {F.}~\bibnamefont
  {Cornet-Gomez}}, \ and\ \bibinfo {author} {\bibfnamefont {M.}~\bibnamefont
  {Nebot}},\ }\href {\doibase 10.1103/PhysRevD.102.035023} {\bibfield
  {journal} {\bibinfo  {journal} {Phys. Rev. D}\ }\textbf {\bibinfo {volume}
  {102}},\ \bibinfo {pages} {035023} (\bibinfo {year} {2020})},\ \Eprint
  {http://arxiv.org/abs/2006.01934} {arXiv:2006.01934 [hep-ph]} \BibitemShut
  {NoStop}%
\bibitem [{\citenamefont {Jung}\ \emph {et~al.}(2010)\citenamefont {Jung},
  \citenamefont {Pich},\ and\ \citenamefont {Tuzon}}]{Jung:2010ik}%
  \BibitemOpen
  \bibfield  {author} {\bibinfo {author} {\bibfnamefont {M.}~\bibnamefont
  {Jung}}, \bibinfo {author} {\bibfnamefont {A.}~\bibnamefont {Pich}}, \ and\
  \bibinfo {author} {\bibfnamefont {P.}~\bibnamefont {Tuzon}},\ }\href
  {\doibase 10.1007/JHEP11(2010)003} {\bibfield  {journal} {\bibinfo  {journal}
  {JHEP}\ }\textbf {\bibinfo {volume} {11}},\ \bibinfo {pages} {003} (\bibinfo
  {year} {2010})},\ \Eprint {http://arxiv.org/abs/1006.0470} {arXiv:1006.0470
  [hep-ph]} \BibitemShut {NoStop}%
\bibitem [{\citenamefont {Li}\ \emph {et~al.}(2014)\citenamefont {Li},
  \citenamefont {Lu},\ and\ \citenamefont {Pich}}]{Li:2014fea}%
  \BibitemOpen
  \bibfield  {author} {\bibinfo {author} {\bibfnamefont {X.-Q.}\ \bibnamefont
  {Li}}, \bibinfo {author} {\bibfnamefont {J.}~\bibnamefont {Lu}}, \ and\
  \bibinfo {author} {\bibfnamefont {A.}~\bibnamefont {Pich}},\ }\href {\doibase
  10.1007/JHEP06(2014)022} {\bibfield  {journal} {\bibinfo  {journal} {JHEP}\
  }\textbf {\bibinfo {volume} {06}},\ \bibinfo {pages} {022} (\bibinfo {year}
  {2014})},\ \Eprint {http://arxiv.org/abs/1404.5865} {arXiv:1404.5865
  [hep-ph]} \BibitemShut {NoStop}%
\bibitem [{\citenamefont {Gunion}\ \emph {et~al.}(2000)\citenamefont {Gunion},
  \citenamefont {Haber}, \citenamefont {Kane},\ and\ \citenamefont
  {Dawson}}]{Gunion:1989we}%
  \BibitemOpen
  \bibfield  {author} {\bibinfo {author} {\bibfnamefont {J.~F.}\ \bibnamefont
  {Gunion}}, \bibinfo {author} {\bibfnamefont {H.~E.}\ \bibnamefont {Haber}},
  \bibinfo {author} {\bibfnamefont {G.~L.}\ \bibnamefont {Kane}}, \ and\
  \bibinfo {author} {\bibfnamefont {S.}~\bibnamefont {Dawson}},\ }\href@noop {}
  {\emph {\bibinfo {title} {{The Higgs Hunter's Guide}}}},\ Vol.~\bibinfo
  {volume} {80}\ (\bibinfo {year} {2000})\BibitemShut {NoStop}%
\bibitem [{\citenamefont {Gunion}\ \emph {et~al.}(1992)\citenamefont {Gunion},
  \citenamefont {Haber}, \citenamefont {Kane},\ and\ \citenamefont
  {Dawson}}]{Gunion:1992hs}%
  \BibitemOpen
  \bibfield  {author} {\bibinfo {author} {\bibfnamefont {J.~F.}\ \bibnamefont
  {Gunion}}, \bibinfo {author} {\bibfnamefont {H.~E.}\ \bibnamefont {Haber}},
  \bibinfo {author} {\bibfnamefont {G.~L.}\ \bibnamefont {Kane}}, \ and\
  \bibinfo {author} {\bibfnamefont {S.}~\bibnamefont {Dawson}},\ }\href@noop {}
  {\  (\bibinfo {year} {1992})},\ \Eprint {http://arxiv.org/abs/hep-ph/9302272}
  {arXiv:hep-ph/9302272} \BibitemShut {NoStop}%
\bibitem [{\citenamefont {Branco}\ \emph {et~al.}(2012)\citenamefont {Branco},
  \citenamefont {Ferreira}, \citenamefont {Lavoura}, \citenamefont {Rebelo},
  \citenamefont {Sher},\ and\ \citenamefont {Silva}}]{Branco:2011iw}%
  \BibitemOpen
  \bibfield  {author} {\bibinfo {author} {\bibfnamefont {G.}~\bibnamefont
  {Branco}}, \bibinfo {author} {\bibfnamefont {P.}~\bibnamefont {Ferreira}},
  \bibinfo {author} {\bibfnamefont {L.}~\bibnamefont {Lavoura}}, \bibinfo
  {author} {\bibfnamefont {M.}~\bibnamefont {Rebelo}}, \bibinfo {author}
  {\bibfnamefont {M.}~\bibnamefont {Sher}}, \ and\ \bibinfo {author}
  {\bibfnamefont {J.~P.}\ \bibnamefont {Silva}},\ }\href {\doibase
  10.1016/j.physrep.2012.02.002} {\bibfield  {journal} {\bibinfo  {journal}
  {Phys. Rept.}\ }\textbf {\bibinfo {volume} {516}},\ \bibinfo {pages} {1}
  (\bibinfo {year} {2012})},\ \Eprint {http://arxiv.org/abs/1106.0034}
  {arXiv:1106.0034 [hep-ph]} \BibitemShut {NoStop}%
\bibitem [{\citenamefont {Delle~Rose}\ \emph
  {et~al.}(2020{\natexlab{b}})\citenamefont {Delle~Rose}, \citenamefont
  {Khalil}, \citenamefont {King},\ and\ \citenamefont
  {Moretti}}]{DelleRose:2019ukt}%
  \BibitemOpen
  \bibfield  {author} {\bibinfo {author} {\bibfnamefont {L.}~\bibnamefont
  {Delle~Rose}}, \bibinfo {author} {\bibfnamefont {S.}~\bibnamefont {Khalil}},
  \bibinfo {author} {\bibfnamefont {S.~J.}\ \bibnamefont {King}}, \ and\
  \bibinfo {author} {\bibfnamefont {S.}~\bibnamefont {Moretti}},\ }\href
  {\doibase 10.1103/PhysRevD.101.115009} {\bibfield  {journal} {\bibinfo
  {journal} {Phys. Rev. D}\ }\textbf {\bibinfo {volume} {101}},\ \bibinfo
  {pages} {115009} (\bibinfo {year} {2020}{\natexlab{b}})},\ \Eprint
  {http://arxiv.org/abs/1903.11146} {arXiv:1903.11146 [hep-ph]} \BibitemShut
  {NoStop}%
\bibitem [{\citenamefont {Barr}\ and\ \citenamefont {Zee}(1990)}]{Barr:1990vd}%
  \BibitemOpen
  \bibfield  {author} {\bibinfo {author} {\bibfnamefont {S.~M.}\ \bibnamefont
  {Barr}}\ and\ \bibinfo {author} {\bibfnamefont {A.}~\bibnamefont {Zee}},\
  }\href {\doibase 10.1103/PhysRevLett.65.21} {\bibfield  {journal} {\bibinfo
  {journal} {Phys. Rev. Lett.}\ }\textbf {\bibinfo {volume} {65}},\ \bibinfo
  {pages} {21} (\bibinfo {year} {1990})},\ \bibinfo {note} {[Erratum:
  Phys.Rev.Lett. 65, 2920 (1990)]}\BibitemShut {NoStop}%
\bibitem [{\citenamefont {Czarnecki}\ \emph {et~al.}(1995)\citenamefont
  {Czarnecki}, \citenamefont {Krause},\ and\ \citenamefont
  {Marciano}}]{Czarnecki:1995wq}%
  \BibitemOpen
  \bibfield  {author} {\bibinfo {author} {\bibfnamefont {A.}~\bibnamefont
  {Czarnecki}}, \bibinfo {author} {\bibfnamefont {B.}~\bibnamefont {Krause}}, \
  and\ \bibinfo {author} {\bibfnamefont {W.~J.}\ \bibnamefont {Marciano}},\
  }\href {\doibase 10.1103/PhysRevD.52.R2619} {\bibfield  {journal} {\bibinfo
  {journal} {Phys. Rev. D}\ }\textbf {\bibinfo {volume} {52}},\ \bibinfo
  {pages} {2619} (\bibinfo {year} {1995})},\ \Eprint
  {http://arxiv.org/abs/hep-ph/9506256} {arXiv:hep-ph/9506256} \BibitemShut
  {NoStop}%
\bibitem [{\citenamefont {Chang}\ \emph {et~al.}(1991)\citenamefont {Chang},
  \citenamefont {Keung},\ and\ \citenamefont {Yuan}}]{Chang:1990sf}%
  \BibitemOpen
  \bibfield  {author} {\bibinfo {author} {\bibfnamefont {D.}~\bibnamefont
  {Chang}}, \bibinfo {author} {\bibfnamefont {W.-Y.}\ \bibnamefont {Keung}}, \
  and\ \bibinfo {author} {\bibfnamefont {T.}~\bibnamefont {Yuan}},\ }\href
  {\doibase 10.1103/PhysRevD.43.R14} {\bibfield  {journal} {\bibinfo  {journal}
  {Phys. Rev. D}\ }\textbf {\bibinfo {volume} {43}},\ \bibinfo {pages} {14}
  (\bibinfo {year} {1991})}\BibitemShut {NoStop}%
\bibitem [{\citenamefont {Cheung}\ \emph {et~al.}(2001)\citenamefont {Cheung},
  \citenamefont {Chou},\ and\ \citenamefont {Kong}}]{Cheung:2001hz}%
  \BibitemOpen
  \bibfield  {author} {\bibinfo {author} {\bibfnamefont {K.-m.}\ \bibnamefont
  {Cheung}}, \bibinfo {author} {\bibfnamefont {C.-H.}\ \bibnamefont {Chou}}, \
  and\ \bibinfo {author} {\bibfnamefont {O.~C.}\ \bibnamefont {Kong}},\ }\href
  {\doibase 10.1103/PhysRevD.64.111301} {\bibfield  {journal} {\bibinfo
  {journal} {Phys. Rev. D}\ }\textbf {\bibinfo {volume} {64}},\ \bibinfo
  {pages} {111301} (\bibinfo {year} {2001})},\ \Eprint
  {http://arxiv.org/abs/hep-ph/0103183} {arXiv:hep-ph/0103183} \BibitemShut
  {NoStop}%
\bibitem [{\citenamefont {Cheung}\ \emph {et~al.}(2009)\citenamefont {Cheung},
  \citenamefont {Kong},\ and\ \citenamefont {Lee}}]{Cheung:2009fc}%
  \BibitemOpen
  \bibfield  {author} {\bibinfo {author} {\bibfnamefont {K.}~\bibnamefont
  {Cheung}}, \bibinfo {author} {\bibfnamefont {O.~C.}\ \bibnamefont {Kong}}, \
  and\ \bibinfo {author} {\bibfnamefont {J.~S.}\ \bibnamefont {Lee}},\ }\href
  {\doibase 10.1088/1126-6708/2009/06/020} {\bibfield  {journal} {\bibinfo
  {journal} {JHEP}\ }\textbf {\bibinfo {volume} {06}},\ \bibinfo {pages} {020}
  (\bibinfo {year} {2009})},\ \Eprint {http://arxiv.org/abs/0904.4352}
  {arXiv:0904.4352 [hep-ph]} \BibitemShut {NoStop}%
\bibitem [{\citenamefont {Cherchiglia}\ \emph {et~al.}(2017)\citenamefont
  {Cherchiglia}, \citenamefont {Kneschke}, \citenamefont {St\"ockinger},\ and\
  \citenamefont {St\"ockinger-Kim}}]{Cherchiglia:2016eui}%
  \BibitemOpen
  \bibfield  {author} {\bibinfo {author} {\bibfnamefont {A.}~\bibnamefont
  {Cherchiglia}}, \bibinfo {author} {\bibfnamefont {P.}~\bibnamefont
  {Kneschke}}, \bibinfo {author} {\bibfnamefont {D.}~\bibnamefont
  {St\"ockinger}}, \ and\ \bibinfo {author} {\bibfnamefont {H.}~\bibnamefont
  {St\"ockinger-Kim}},\ }\href {\doibase 10.1007/JHEP01(2017)007} {\bibfield
  {journal} {\bibinfo  {journal} {JHEP}\ }\textbf {\bibinfo {volume} {01}},\
  \bibinfo {pages} {007} (\bibinfo {year} {2017})},\ \Eprint
  {http://arxiv.org/abs/1607.06292} {arXiv:1607.06292 [hep-ph]} \BibitemShut
  {NoStop}%
\bibitem [{\citenamefont {Pe\~nuelas}\ and\ \citenamefont
  {Pich}(2017)}]{Penuelas:2017ikk}%
  \BibitemOpen
  \bibfield  {author} {\bibinfo {author} {\bibfnamefont {A.}~\bibnamefont
  {Pe\~nuelas}}\ and\ \bibinfo {author} {\bibfnamefont {A.}~\bibnamefont
  {Pich}},\ }\href {\doibase 10.1007/JHEP12(2017)084} {\bibfield  {journal}
  {\bibinfo  {journal} {JHEP}\ }\textbf {\bibinfo {volume} {12}},\ \bibinfo
  {pages} {084} (\bibinfo {year} {2017})},\ \Eprint
  {http://arxiv.org/abs/1710.02040} {arXiv:1710.02040 [hep-ph]} \BibitemShut
  {NoStop}%
\bibitem [{\citenamefont {Botella}\ \emph {et~al.}(2018)\citenamefont
  {Botella}, \citenamefont {Cornet-Gomez},\ and\ \citenamefont
  {Nebot}}]{Botella:2018gzy}%
  \BibitemOpen
  \bibfield  {author} {\bibinfo {author} {\bibfnamefont {F.~J.}\ \bibnamefont
  {Botella}}, \bibinfo {author} {\bibfnamefont {F.}~\bibnamefont
  {Cornet-Gomez}}, \ and\ \bibinfo {author} {\bibfnamefont {M.}~\bibnamefont
  {Nebot}},\ }\href {\doibase 10.1103/PhysRevD.98.035046} {\bibfield  {journal}
  {\bibinfo  {journal} {Phys. Rev. D}\ }\textbf {\bibinfo {volume} {98}},\
  \bibinfo {pages} {035046} (\bibinfo {year} {2018})},\ \Eprint
  {http://arxiv.org/abs/1803.08521} {arXiv:1803.08521 [hep-ph]} \BibitemShut
  {NoStop}%
\bibitem [{\citenamefont {Abbiendi}\ \emph {et~al.}(2013)\citenamefont
  {Abbiendi} \emph {et~al.}}]{Abbiendi:2013hk}%
  \BibitemOpen
  \bibfield  {author} {\bibinfo {author} {\bibfnamefont {G.}~\bibnamefont
  {Abbiendi}} \emph {et~al.} (\bibinfo {collaboration} {ALEPH, DELPHI, L3,
  OPAL, LEP}),\ }\href {\doibase 10.1140/epjc/s10052-013-2463-1} {\bibfield
  {journal} {\bibinfo  {journal} {Eur. Phys. J. C}\ }\textbf {\bibinfo {volume}
  {73}},\ \bibinfo {pages} {2463} (\bibinfo {year} {2013})},\ \Eprint
  {http://arxiv.org/abs/1301.6065} {arXiv:1301.6065 [hep-ex]} \BibitemShut
  {NoStop}%
\bibitem [{\citenamefont {Bechtle}\ \emph {et~al.}(2014)\citenamefont
  {Bechtle}, \citenamefont {Heinemeyer}, \citenamefont {St\r{a}l},
  \citenamefont {Stefaniak},\ and\ \citenamefont {Weiglein}}]{Bechtle:2013xfa}%
  \BibitemOpen
  \bibfield  {author} {\bibinfo {author} {\bibfnamefont {P.}~\bibnamefont
  {Bechtle}}, \bibinfo {author} {\bibfnamefont {S.}~\bibnamefont {Heinemeyer}},
  \bibinfo {author} {\bibfnamefont {O.}~\bibnamefont {St\r{a}l}}, \bibinfo
  {author} {\bibfnamefont {T.}~\bibnamefont {Stefaniak}}, \ and\ \bibinfo
  {author} {\bibfnamefont {G.}~\bibnamefont {Weiglein}},\ }\href {\doibase
  10.1140/epjc/s10052-013-2711-4} {\bibfield  {journal} {\bibinfo  {journal}
  {Eur. Phys. J. C}\ }\textbf {\bibinfo {volume} {74}},\ \bibinfo {pages}
  {2711} (\bibinfo {year} {2014})},\ \Eprint {http://arxiv.org/abs/1305.1933}
  {arXiv:1305.1933 [hep-ph]} \BibitemShut {NoStop}%
\bibitem [{\citenamefont {Bechtle}\ \emph {et~al.}(2020)\citenamefont
  {Bechtle}, \citenamefont {Dercks}, \citenamefont {Heinemeyer}, \citenamefont
  {Klingl}, \citenamefont {Stefaniak}, \citenamefont {Weiglein},\ and\
  \citenamefont {Wittbrodt}}]{Bechtle:2020pkv}%
  \BibitemOpen
  \bibfield  {author} {\bibinfo {author} {\bibfnamefont {P.}~\bibnamefont
  {Bechtle}}, \bibinfo {author} {\bibfnamefont {D.}~\bibnamefont {Dercks}},
  \bibinfo {author} {\bibfnamefont {S.}~\bibnamefont {Heinemeyer}}, \bibinfo
  {author} {\bibfnamefont {T.}~\bibnamefont {Klingl}}, \bibinfo {author}
  {\bibfnamefont {T.}~\bibnamefont {Stefaniak}}, \bibinfo {author}
  {\bibfnamefont {G.}~\bibnamefont {Weiglein}}, \ and\ \bibinfo {author}
  {\bibfnamefont {J.}~\bibnamefont {Wittbrodt}},\ }\href@noop {} {\  (\bibinfo
  {year} {2020})},\ \Eprint {http://arxiv.org/abs/2006.06007} {arXiv:2006.06007
  [hep-ph]} \BibitemShut {NoStop}%
\bibitem [{\citenamefont {Kuno}\ and\ \citenamefont
  {Okada}(2001)}]{Kuno:1999jp}%
  \BibitemOpen
  \bibfield  {author} {\bibinfo {author} {\bibfnamefont {Y.}~\bibnamefont
  {Kuno}}\ and\ \bibinfo {author} {\bibfnamefont {Y.}~\bibnamefont {Okada}},\
  }\href {\doibase 10.1103/RevModPhys.73.151} {\bibfield  {journal} {\bibinfo
  {journal} {Rev. Mod. Phys.}\ }\textbf {\bibinfo {volume} {73}},\ \bibinfo
  {pages} {151} (\bibinfo {year} {2001})},\ \Eprint
  {http://arxiv.org/abs/hep-ph/9909265} {arXiv:hep-ph/9909265} \BibitemShut
  {NoStop}%
\bibitem [{\citenamefont {Abe}\ \emph {et~al.}(2015)\citenamefont {Abe},
  \citenamefont {Sato},\ and\ \citenamefont {Yagyu}}]{Abe:2015oca}%
  \BibitemOpen
  \bibfield  {author} {\bibinfo {author} {\bibfnamefont {T.}~\bibnamefont
  {Abe}}, \bibinfo {author} {\bibfnamefont {R.}~\bibnamefont {Sato}}, \ and\
  \bibinfo {author} {\bibfnamefont {K.}~\bibnamefont {Yagyu}},\ }\href
  {\doibase 10.1007/JHEP07(2015)064} {\bibfield  {journal} {\bibinfo  {journal}
  {JHEP}\ }\textbf {\bibinfo {volume} {07}},\ \bibinfo {pages} {064} (\bibinfo
  {year} {2015})},\ \Eprint {http://arxiv.org/abs/1504.07059} {arXiv:1504.07059
  [hep-ph]} \BibitemShut {NoStop}%
\bibitem [{\citenamefont {Zyla}\ \emph {et~al.}(2020)\citenamefont {Zyla} \emph
  {et~al.}}]{Zyla:2020zbs}%
  \BibitemOpen
  \bibfield  {author} {\bibinfo {author} {\bibfnamefont {P.}~\bibnamefont
  {Zyla}} \emph {et~al.} (\bibinfo {collaboration} {Particle Data Group}),\
  }\href {\doibase 10.1093/ptep/ptaa104} {\bibfield  {journal} {\bibinfo
  {journal} {PTEP}\ }\textbf {\bibinfo {volume} {2020}},\ \bibinfo {pages}
  {083C01} (\bibinfo {year} {2020})}\BibitemShut {NoStop}%
\bibitem [{\citenamefont {Alwall}\ \emph {et~al.}(2014)\citenamefont {Alwall},
  \citenamefont {Frederix}, \citenamefont {Frixione}, \citenamefont {Hirschi},
  \citenamefont {Maltoni}, \citenamefont {Mattelaer}, \citenamefont {Shao},
  \citenamefont {Stelzer}, \citenamefont {Torrielli},\ and\ \citenamefont
  {Zaro}}]{Alwall:2014hca}%
  \BibitemOpen
  \bibfield  {author} {\bibinfo {author} {\bibfnamefont {J.}~\bibnamefont
  {Alwall}}, \bibinfo {author} {\bibfnamefont {R.}~\bibnamefont {Frederix}},
  \bibinfo {author} {\bibfnamefont {S.}~\bibnamefont {Frixione}}, \bibinfo
  {author} {\bibfnamefont {V.}~\bibnamefont {Hirschi}}, \bibinfo {author}
  {\bibfnamefont {F.}~\bibnamefont {Maltoni}}, \bibinfo {author} {\bibfnamefont
  {O.}~\bibnamefont {Mattelaer}}, \bibinfo {author} {\bibfnamefont {H.~S.}\
  \bibnamefont {Shao}}, \bibinfo {author} {\bibfnamefont {T.}~\bibnamefont
  {Stelzer}}, \bibinfo {author} {\bibfnamefont {P.}~\bibnamefont {Torrielli}},
  \ and\ \bibinfo {author} {\bibfnamefont {M.}~\bibnamefont {Zaro}},\ }\href
  {\doibase 10.1007/JHEP07(2014)079} {\bibfield  {journal} {\bibinfo  {journal}
  {JHEP}\ }\textbf {\bibinfo {volume} {07}},\ \bibinfo {pages} {079} (\bibinfo
  {year} {2014})},\ \Eprint {http://arxiv.org/abs/1405.0301} {arXiv:1405.0301
  [hep-ph]} \BibitemShut {NoStop}%
\end{thebibliography}%

\end{document}